\documentclass[fleqn,usenatbib]{mnras}

\usepackage{newtxtext,newtxmath}
\usepackage[T1]{fontenc}
\usepackage{ae,aecompl}

\usepackage{graphicx}	% Including figure files
\usepackage{amsmath}	% Advanced maths commands
\title[\textit{An X-ray and radio study of AS1063}]{An X-ray and Radio Study of the Hubble Frontier Field Cluster Abell S1063}

\author[Rahaman et al.]{
Majidul Rahaman$^{1}$\thanks{E-mail: phd1601121007@iiti.ac.in},
Ramij Raja$^{1}$,
Abhirup Datta$^{1}$,
Jack O Burns$^{2}$,
Brian Alden$^{2}$,
\newauthor
and David Rapetti$^{3,4,2}$ \\
$^{1}$\href{http://astronomy.iiti.ac.in/index.html}{Discipline of Astronomy, Astrophysics and Space Engineering}, \href{http://www.iiti.ac.in/}{Indian Institute Of Technology Indore}, Indore, India\\
$^{2}$Center for Astrophysics and Space Astronomy, Department of Astrophysical \& Planterary Sciences, University of Colorado, Boulder,\\ CO 80309, USA\\
$^{3}$NASA Ames Research Center, Moffett Field, CA 94035, USA\\
$^{4}$Research Institute for Advanced Computer Science, Universities Space Research Association, Mountain View, CA 94043, USA \\
}
\date{Accepted 2021 April 26. Received 2021 April 23; in original form 2020 August 21}
\pubyear{2021}
\begin{document}
\label{firstpage}
\pagerange{\pageref{firstpage}--\pageref{lastpage}}
\maketitle

\begin{abstract}
We present results from \textit{Chandra} X-ray observations and 325 MHz Giant Metrewave Radio Telescope (GMRT) observations of the massive and X-ray luminous cluster of galaxies Abell S1063.
We report the detection of large-scale \lq\lq excess brightness\rq\rq\ in the residual \textit{Chandra} X-ray surface brightness map, which extends at least 2.7 Mpc towards the north-east from the center of the cluster.
We also present a high fidelity X-ray flux and temperature map using \textit{Chandra} archival data of 122 ksec, which shows the disturbed morphology in the cluster. The residual flux map shows the first observational confirmation of the merging axis proposed by earlier simulation by \citet{Gomez2012AJ....144...79G}.
The average temperature within $R_{500}$ is  $11.7 \pm 0.56$ keV,  which makes AS1063 one of the hottest clusters in the nearby Universe..
The integrated radio flux density at 325 MHz is found to be $62.0\pm6.3$ mJy.
The integrated spectrum of the radio halo follows a power-law with a spectral index $\alpha=-1.43\pm 0.13$.  The radio halo is found to be significantly under-luminous, which favored for both the hadronic as well as the turbulent re-acceleration mechanism for its origin.
\end{abstract}


%%%%%%%%%%%%%%%%%%%%%%%%%%%%%%%%%%%%%%%%%%%%%%%%%%

%%%%%%%%%%%%%%%%% BODY OF PAPER %%%%%%%%%%%%%%%%%%

\section{Introduction} \label{intro}
Galaxy clusters are assembled through major and minor mergers. Major mergers are the most energetic events after the Big Bang and release as much as $10^{64}$ ergs of energy within a timescale of the order of a few Gyrs \citep{Gastaldello2003A&A...411...21G}.
With the help of high-resolution X-ray telescopes (e.g., \textit{Chandra, XMM}-Newton), observational evidence of merger-induced shocks in the intracluster medium (ICM) is common in surface brightness maps (or residuals of surface brightness maps), and temperature maps \citep{Hallman2018}.
Mpc scale diffuse synchrotron radio emission is sometimes found in high-mass ($\mathrm{M_{500}} \geq 5 \times 10^{14} \mathrm{M_{\odot}}$) merging clusters and is not directly connected with any cluster radio galaxy \citep{Cassano2019arXiv190710304C,van_Weeren2019SSRv..215...16V}.
If this emission fills the central Mpc region of the host cluster, then it is called a radio halo (RH). 

RHs are most likely formed during cluster mergers in which a considerable amount of energy is injected into the ICM, which causes turbulent motions and shocks.
RHs are produced by ultra-relativistic electrons with Lorentz factors $\gamma \sim 10^{3}-10^{4}$ in the presence of large-scale $\sim\mu$G magnetic fields.
The physical mechanism powering this emission has been historically debated between two models: hadronic \citep{Dennison1980ApJ...239L..93D,Blasi1999APh....12..169B,Dolag2000A&A...362..151D,Miniati2001ApJ...559...59M,Miniati2001ApJ...562..233M,Pfrommer2008MNRAS.385.1211P,Enblin2011A&A...527A..99E} and turbulent re-acceleration \citep{Brunetti2001MNRAS.320..365B,Petrosian2001ApJ...557..560P,Donnert2013MNRAS.429.3564D}.

According to the hadronic model, RHs result from the relativistic electrons produced in inelastic hadronic interactions between cosmic ray protons ($\mathrm{CR_p}$) and thermal protons. 
There are some limitations with the hadronic model.
It predicts $\gamma$-ray emission that has not been observed in nearby galaxy clusters by the \textit{Fermi} satellite \citep{Ackermann2014ApJ...787...18A,Brunetti_2017MNRAS.472.1506B}.
Further, the hadronic model cannot explain RHs with very steep spectral indices  ($\alpha < -1.5$, as $S_\nu \propto \nu^{\alpha}$  \citealt{Brunetti2008Natur.455..944B}). The model proposed in  \citet{Keshet2010ApJ...722..737K} requires different magnetic fields for the galaxy clusters that host a RH and those without a RH. This has never been confirmed by radio observations \citep{Bonafede2011A&A...530A..24B,Govoni2010A&A...522A.105G}.
Furthermore, the hadronic model cannot explain the correlation between X-ray luminosity ($L_X$) and radio power ($P_{1.4})$.
\citet{Brunetti2011MNRAS.410..127B} showed that electrons originating from hadronic collisions could generate RHs a factor of $\sim$10 below the $L_{X}-P_{1.4\ GHz}$ correlation. 

In the re-acceleration model, an existing population of seed relativistic electrons are re-accelerated during a powerful state of ICM turbulence due to cluster-mergers events, which produces synchrotron emission in the presence of magnetic fields \citep{Brunetti2001MNRAS.320..365B,Petrosian2001ApJ...557..560P,Donnert2013MNRAS.429.3564D}.
An open question about the re-acceleration model is the source of these seed electrons,
which may be coming from primary electrons from supernovae, active galactic nuclei, galaxies, and shocks in the cluster, which can be accumulated for a few Gyr at energies of a few hundred MeV \citep{Sarazin1999ApJ...520..529S,Brunetti2001MNRAS.320..365B}.
Turbulence re-acceleration is believed to be the primary physical mechanism by which RHs are formed.

There are other mechanisms that have been proposed to explain the origin of RHs such as: the hybrid model (a mixture of the hadronic and re-acceleration models) \citep{Brunetti2011MNRAS.410..127B,Zandanel2014MNRAS.438..124Z}, and the magnetic reconnection model \citep{Brunetti2016MNRAS.458.2584B}. In addition, 
\citet{Cassano2010ApJ...721L..82C} showed that RHs are associated with dynamically disturbed clusters, and clusters without RHs are more relaxed, with only a few exceptions where a disturbed cluster does not exhibit a RH.

Here, we present the case of Abell S1063 (\citealt{Abell1989ApJS...70....1A}, hereafter AS1063, also known as SPT-CL J2248-4431;  RXC J2248.7-4431 ; PLCKESZ G349.46-59.94), which is a Hubble Frontier Field cluster in an early-stage of merging \citep{Gomez2012AJ....144...79G}. It has been studied before in the optical \citep{Gomez2012AJ....144...79G,Ruel2014ApJ...792...45R,Saro2017MNRAS.468.3347S}, X-ray \citep{Maughan2008ApJS..174..117M,Gomez2012AJ....144...79G,McDonald2013ApJ...774...23M,Shitanishi2018MNRAS.481..749S}, and via weak lensing \citep{Gruen2013MNRAS.432.1455G}. 
It is a massive cluster with $M^{\mathrm{SZ}}_{500} = 17.97 \pm 2.18 \times 10^{14} M_\odot$ \citep{Birzan2017MNRAS.471.1766B}. 
This massive and bright cluster is going through a major merger event, similar to that of the bullet cluster \citep{Mastropietro_2008}.
Recently, \citet{Xie_2020A&A...636A...3X} discovered a radio halo in this cluster. They found that the integrated spectral index of the radio halo steepens at higher frequencies.
In general, radio halos are present in massive and disturbed clusters \citep{cassano2013ApJ...777..141C,Kale_2015A&A...579A..92K}. 
However, the dynamical state of the AS1063 is still debated. Some studies suggest that this is a relaxed cluster (\citet{McDonald2013ApJ...774...23M,Lovisari2017ApJ...846...51L}) whereas, \citet{Gomez2012AJ....144...79G,Gruen2013MNRAS.432.1455G} reported it as a disturbed cluster.
Since X-ray temperature maps can reveal the dynamical state of a cluster more accurately, using \textit{Chandra} X-ray data here we present a high-resolution temperature map of AS1063 showing the disturbed state of this cluster.

In this paper, we analyze archival \textit{Chandra} X-ray observations and 325 MHz radio observations from the Giant Metre-wave Radio Telescope (GMRT). In section \ref{data_reduction}, we discuss the data analysis: calibration and imaging of both the \textit{Chandra} X-ray and 325 MHz GMRT observations. In section \ref{result}, we present the results from both the X-ray and radio observations. In section \ref{sec:discussion}, we discuss all the results, and conclusions are presented in section \ref{sec:conclude}.

Throughout the paper we assume a $\Lambda$CDM cosmology with $H_0 = 70$ km s$^{-1}$ Mpc$^{-1}$, $\Omega_{m} = 0.3$ and $\Omega_\Lambda = 0.7$. At the cluster redshift $z = 0.351$, $1\arcsec$ corresponds to a physical scale of 4.95 kpc. Errors are quoted at the 1$\sigma$ level unless noted otherwise.

\section{Observations and data Analysis} \label{data_reduction}
\subsection{\textit{Chandra} X-Ray Observations}
\label{sec:X-Ray}
We analyzed archival\footnote{https://cda.harvard.edu/chaser/} data for three separate observations (Obs-IDs 4966 (PI Romer; 2004), 18611, 18818 (PI Kraft; 2016)) of AS1063 with the \textit{Chandra} X-ray observatory. The entire 122 ksec data were taken in VFAINT mode.
For this study, we employed a systematic calibration and analysis pipeline, which uses the \textit{Chandra} Interactive Analysis of Observations (\textit{CIAO}) and subsequent scripts in IDL and python.

The details of our data reduction pipeline are described in \citet{Datta2014,Schenck2014,Hallman2018,Raja_2020ApJ...889..128R}, which initially consisted of several bash and IDL scripts. The new version of the pipeline in python was developed and recently released as {\it ClusterPyXT}\footnote{https://github.com/bcalden/ClusterPyXT} \citep{Alden_2019ascl.soft05022A}.
In this paper, we used the older version of the pipeline in bash and IDL scripts and parallelized parts of the pipeline to make it more efficient.
It takes \textit{Chandra} observation Ids and generates high fidelity adaptive circularly binned (ACB) temperature, pressure, and entropy maps.
Once observational Ids are supplied by the user, the pipeline automatically downloads data from the \textit{Chandra} archive using the \textit{CIAO} task \textit{download\_chandra\_obsid}, and cleans it in a standard manner for both data and background.
\\
As AS1063 is a comparatively low redshift cluster ($z = 0.351$), the extended x-ray emission of the cluster fills chip 3 and spills over the other three ACIS-I chips. Therefore, we were not able to use the local background (as in e.g., \citet{Raja_2020ApJ...889..128R}). Hence, we had to model the background contribution present in the observation by extracting background spectra from the ''blank-sky'' background files. These ''blank-sky'' background files available in the \textit{Chandra} calibration database (CALDB), which represents particle background and unresolved cosmic X-ray background.
After cleaning all data, it merges all ObsIds to create a combined flux map.
 We created light curves for individual ObsId in the full energy band and the 9.5-12 keV band. 
Light curves were binned at 259 seconds per bin for data as well as blank-sky backgrounds, and count rates higher than 3$\sigma$ were removed (background flares) using the \textit{deflare} tool.
Next, we removed point sources from the data by providing \textit{SAOImage DS9} region files containing point sources using the tool {\it wavdetect} inbuilt into \textit{CIAO} in the 0.2-12 keV band with the scales of 1, 2, 4, 8, and 16 pixels. 
Point sources were inspected visually for any false detection or if {\it wavdetect} failed to detect any real sources. 
Regions with point sources were removed from both data and blank-sky background files to avoid negative subtraction. 
These steps produced calibrated and clean data free from bad events as well as contaminating point sources.

In the next section, we proceed towards making X-ray surface brightness and temperature maps.

\subsubsection{X-ray Surface Brightness Map}
\begin{figure}
\centering
\includegraphics[width=\columnwidth]{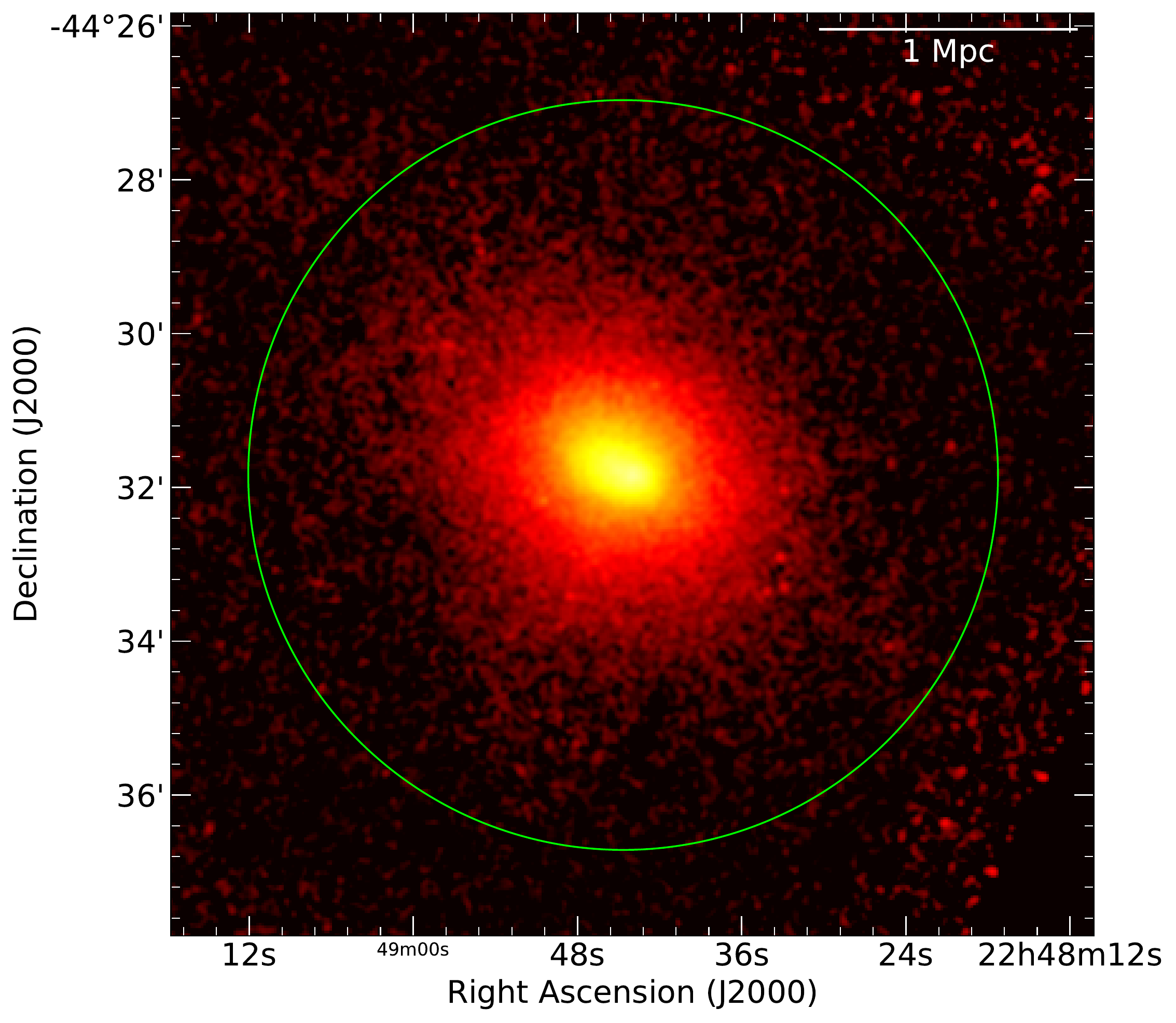}
\caption{Exposure corrected, background subtracted, point source removed, surface brightness map of galaxy cluster AS1063 in 0.7-8.0 keV energy range. The green circle represents R$_{500}$ of the cluster. }
\label{fig:X-Ray-SB}
\end{figure}

After cleaning the data, we combined all data files (3 ObsIds) using \textit{merge\_obs} with binning 4 to produce a surface brightness map. The exposure corrected, background and point sources subtracted, 0.7-8.0 keV surface brightness image is shown in Figure \ref{fig:X-Ray-SB}.

%%%%%%%%%%%%%%%%%%%%%%%%%%%% Figure X-ray SB %%%%%%%%%%%%%%%%%%%%%%

\subsubsection{X-ray Temperature Map} \label{sec:TMAP}
The Adaptive Circular Binning method (ACB, \citealt{Datta2014,Schenck2014}) was used to make a high-resolution temperature map from \textit{Chandra} ACIS-I data. 
Spectra were extracted from circular regions that were large enough to have a signal-to-noise ratio (SNR) of 50.
Here, the signal is background-subtracted clean data, and the noise is Poisson distributed coming from both source and background.
The source and background spectra were extracted using \textit{dmextract} while the weighted response files were extracted using \textit{specextraxt}. 
Since the \textit{Chandra} effective area in the 9.5-12 keV energy range is negligible, almost all of the 9.5-12 keV flux in the sky data is due to particle background. 
So, we re-scaled the background spectra using the ratio of the high energy counts (9.5-12 keV) in the source and the background. One should note that re-scaling in this energy range may introduce $\pm 2\%$ systematic errors \citep{Hickox_2006ApJ...645...95H,Bonamente_2013MNRAS.428.2812B,Bartalucci2014A&A...566A..25B}
APEC and PHABS models were used for spectral fitting for every region in the energy range of 0.7-8.0 keV.
The APEC is a collisional-radiative plasma model that uses atomic data in the companion Astrophysical Plasma Emission Database (APED) to calculate the spectral model for hot plasma \citep{Randall2001ApJ...556L..91S}. PHABS is a photoelectric absorption model. 
The best-fitted temperature (APEC model parameter) and errors (68\% confidence levels using a Markov Chain Monte Carlo algorithm) of each circular spectral regions were assigned to the center of the circle. In this process, circles were allowed to overlap with each other. 
Previously, this method was applied to create
temperature maps for A85 \citep{Schenck2014}, A3667 \citep{Datta2014}, Abell 115 \citep{Hallman2018}, and Phoenix cluster \citep{Raja_2020ApJ...889..128R}.

% Table
\begin{table}
\renewcommand{\arraystretch}{1.5}
\caption{Dynamical state parameters and properties of AS1063.}
\label{tab:parameters}
\begin{tabular}{lcc}
\hline
Parameter & Value & Remark \\
\hline
$M^{\mathrm{SZ}}_{500}$ [$10^{14} \mathrm{M_{\odot}} $] & $17.97 \pm 2.18$ & ...$^{1}$ \\
$\mathrm{t_{cool,0}}$ [Gyr] & $1.79^{+0.14}_{-0.14}$ & $\mathrm{Weak-CC^{2}}$\\
$K_0$ [keV cm$^2$] & $128.0^{+9.5}_{-9.4}$ & $\mathrm{non-CC^{2}}$\\
$\mathrm{dM/dt_{7.7}}\ [\mathrm{M_{\odot}\ yr^{-1}}]$ & $652.4^{+48.2}_{-42.0}$ & $...^{2}$\\
$\mathrm{dM/dt_{Univ}}\ [\mathrm{M_{\odot}\ yr^{-1}}]$ & $997.1^{+73.6}_{-64.1}$ & $...^{2}$\\
$w$ & $0.006^{+0.001}_{-0.000}$ & Stable$^{2}$\\
$\mathrm{A_{phot}}$ & $0.21^{+0.03}_{-0.02}$ & moderately \\
 & & disturbed$^{3}$ \\
 $\mathrm{c_{SB}}$ & $0.23^{+0.01}_{-0.01}$ & $\mathrm{CC}$\\
 $\mathrm{L_{X}^{*}}$ [$ 10^{45}\ \mathrm{ ergs.s^{-1}} $] & $2.51^{+0.02}_{-0.03}$ & \\
\hline
$\mathrm{S_{1.5\ GHz}^{VLA}}$ [Halo] & $5.8\pm0.4 \mathrm{mJy}$ & $...^{4}$\\
$\mathrm{S_{3.0\ GHz}^{VLA}}$ [Halo] & $1.7\pm0.2$ mJy & $...^{4}$\\
\hline
$\mathrm{S_{325\ MHz}^{GMRT}}$ [Halo] & $62.0\pm6.28$ mJy & \\
$\mathrm{P_{1.4}}$ [$10^{+24}$ W/Hz ] & $3.63 \pm 0.37$ & \\
\hline
\end{tabular}
\\ * \textit{Chandra} X-ray luminosity ($\mathrm{L_{X}}$) calculated in the 0.1-2.4 keV energy band.
\\References: 1. \citet{Birzan2017MNRAS.471.1766B} 2.  \citet{McDonald2013ApJ...774...23M}, 3. \citet{Nurgaliev2017ApJ...841....5N}, 4. \citet{Xie_2020A&A...636A...3X}.
\end{table}

We performed spectral fitting using the \textit{XSPEC} version: 12.9.1 in the 0.7-8.0 keV energy range. The APEC and PHABS models are fitted to the spectra of each region. All three spectra (from each obsId) were fitted simultaneously from each ACB region using the C-statistics \citep{Cash1979}. 
The metallicity of the cluster was kept frozen at 0.3 $Z_\odot$,  which can add a systematic to the derived results. We calculated the abundance profile (see Table \ref{tab:abundance_table}1) using the same binning as described in the azimuthal temperature profile in Figure \ref{fig:T_profile}. We also calculated the temperature profile applying the newly derived abundance table and found that the new temperature profile agrees with the previous one within the 1$\sigma$ level (see Table \ref{tab:abundance_table}1). Redshift z = 0.351 and the Galactic hydrogen column density $N_H = 1.24\times10^{20}$ $cm^{-2}$ (the weighted average $N_H$ from the Leiden Argentine Bonn (LAB) survey, \citealt{Kalberla2005A&A...440..775K}) were used for this analysis. Only APEC normalization and temperature parameters were fitted for each spectrum. Temperature and corresponding error maps were calculated within 1$\sigma$ confidence level. The event counts decrease at the periphery of the cluster. Due to the constraint in the signal to noise ratio, we masked beyond 800 kpc in the temperature map.

\begin{figure*}
\centering
\begin{tabular}{lccr}
\includegraphics[width=\columnwidth]{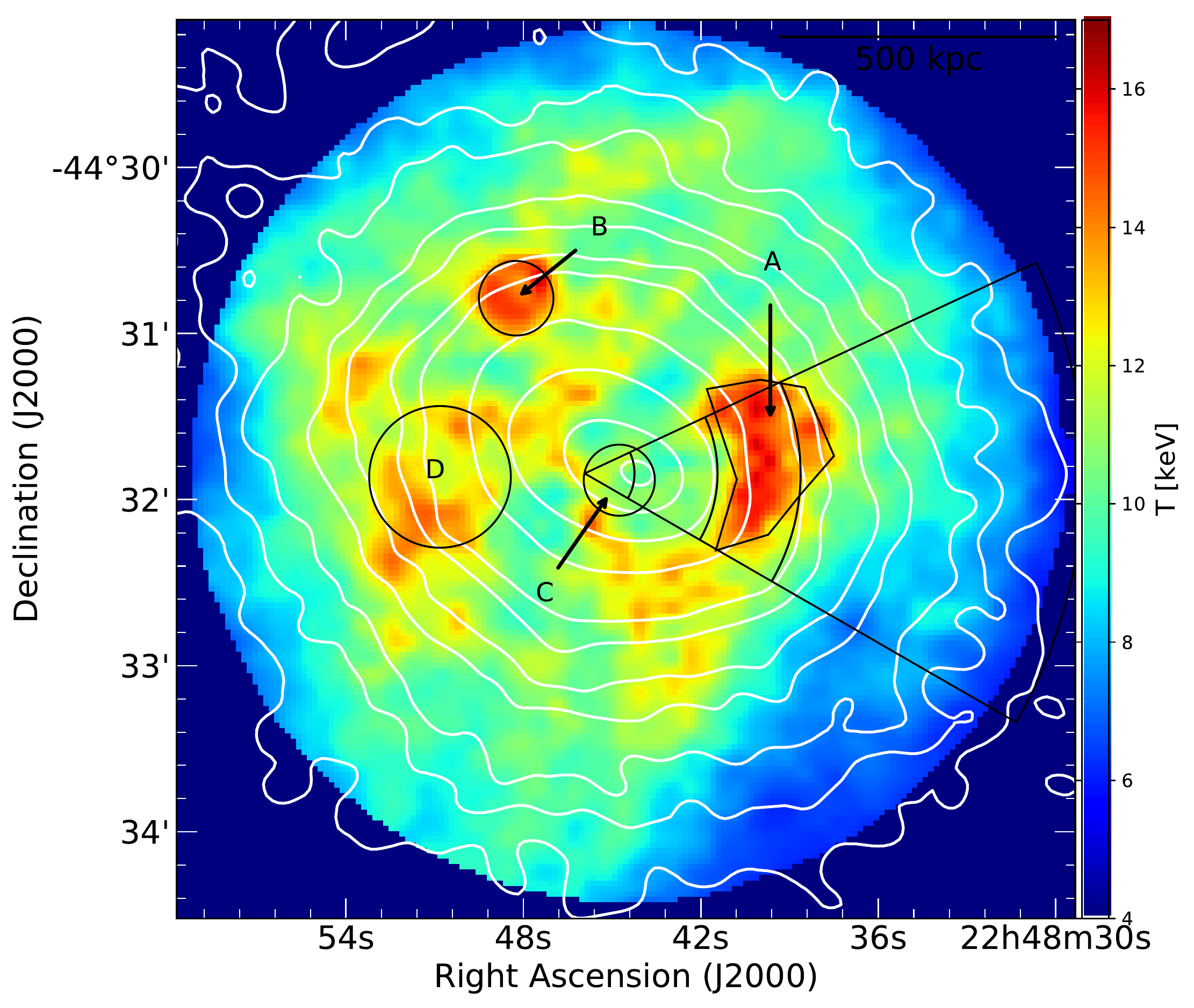} &
\includegraphics[width=\columnwidth]{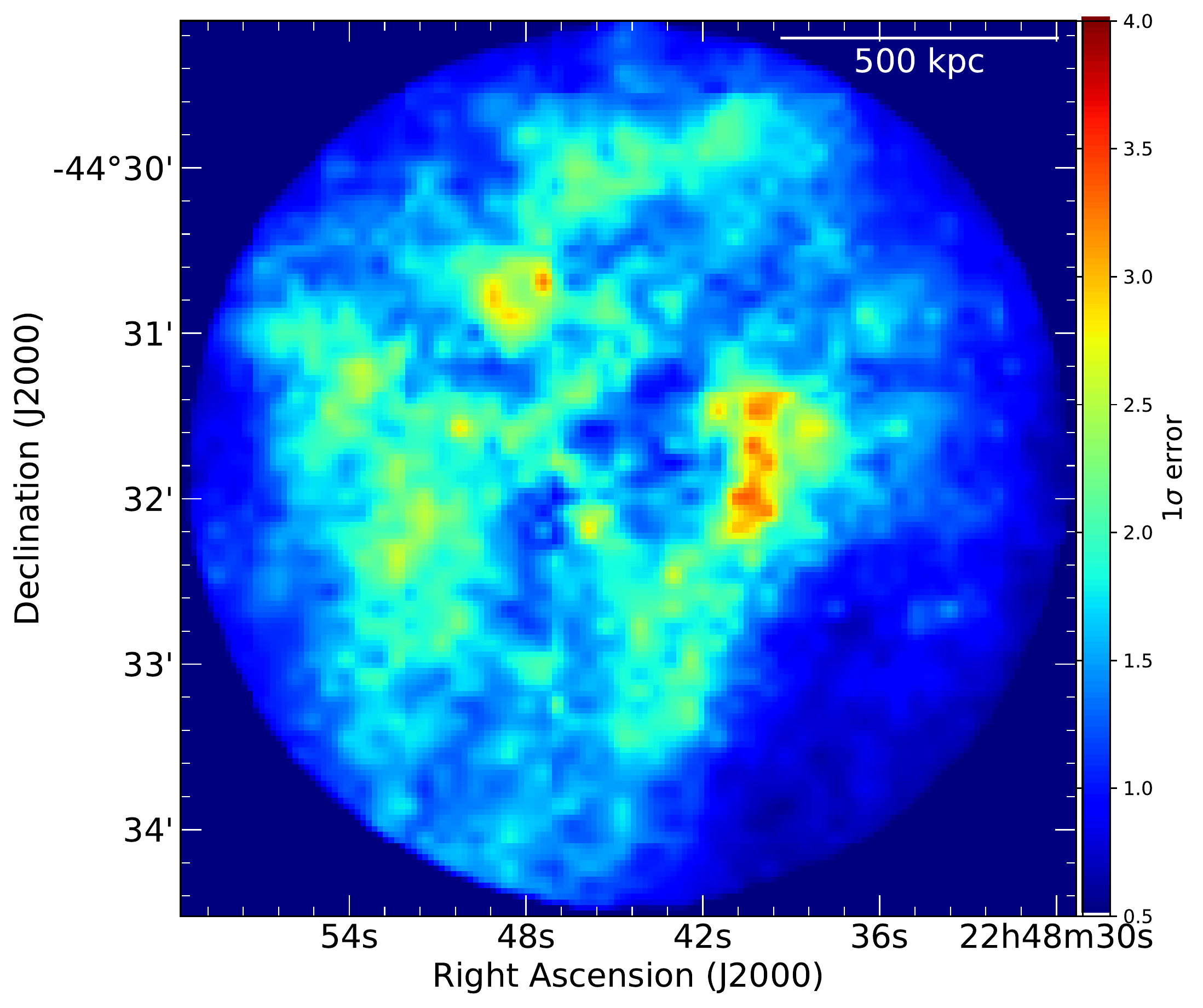} \\

\end{tabular}
\caption{Left: \textit{Chandra} X-ray temperature map of AS1063 constructed using adaptive circular binning (ACB) as described in Section \ref{sec:TMAP}, overlaid with \textit{Chandra} X-ray surface brightness contours (white). Temperature profile along the wedge region is shown in right panel of Figure \ref{fig:T_profile}. Right: 1$\sigma$ error map corresponding to the ACB temperature map on the left. These images are used as a guide to better understand the dynamical state of the cluster. Due to low counts at the outskirts of the cluster, we restrict the map only to the inner 800 kpc of the cluster.}
\label{fig:ACB_Tmap}
\end{figure*}

The ACB temperature map for AS1063, along with the 1$\sigma$ percent error map for the best-fitted temperatures, is shown in Figure \ref{fig:ACB_Tmap}.

%%%%%%%%%%%%%%%%%%%%%%%%%%%% Figure ACB Tmap and Error %%%%%%%%%%%%%%%%%%%%%%%%%

\subsubsection{X-ray imaging analysis} \label{sec:filament}
\begin{figure*}
\centering
\begin{tabular}{cc}
\includegraphics[width=3.3in,height=3.3in]{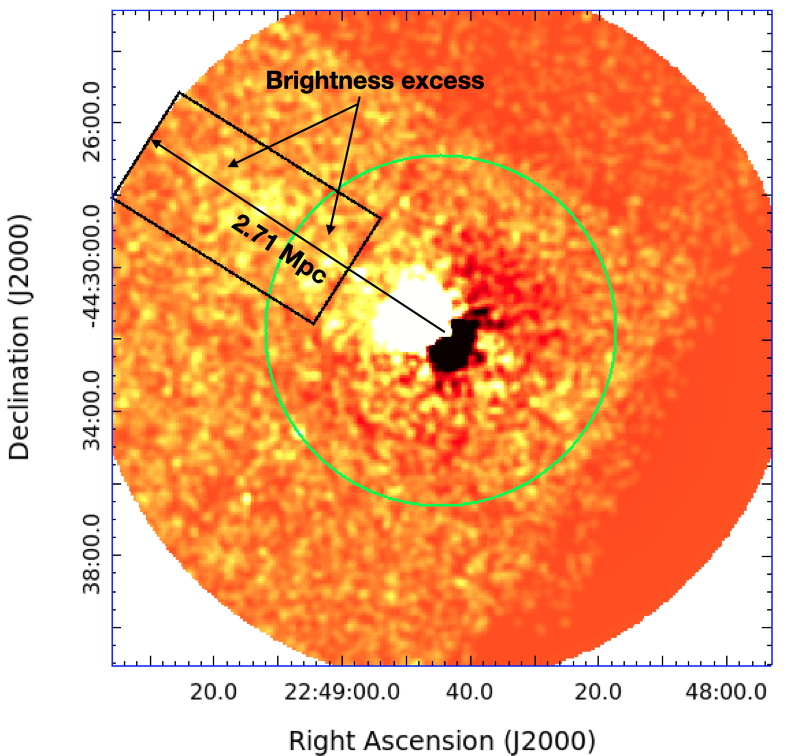} &
\includegraphics[width=\columnwidth]{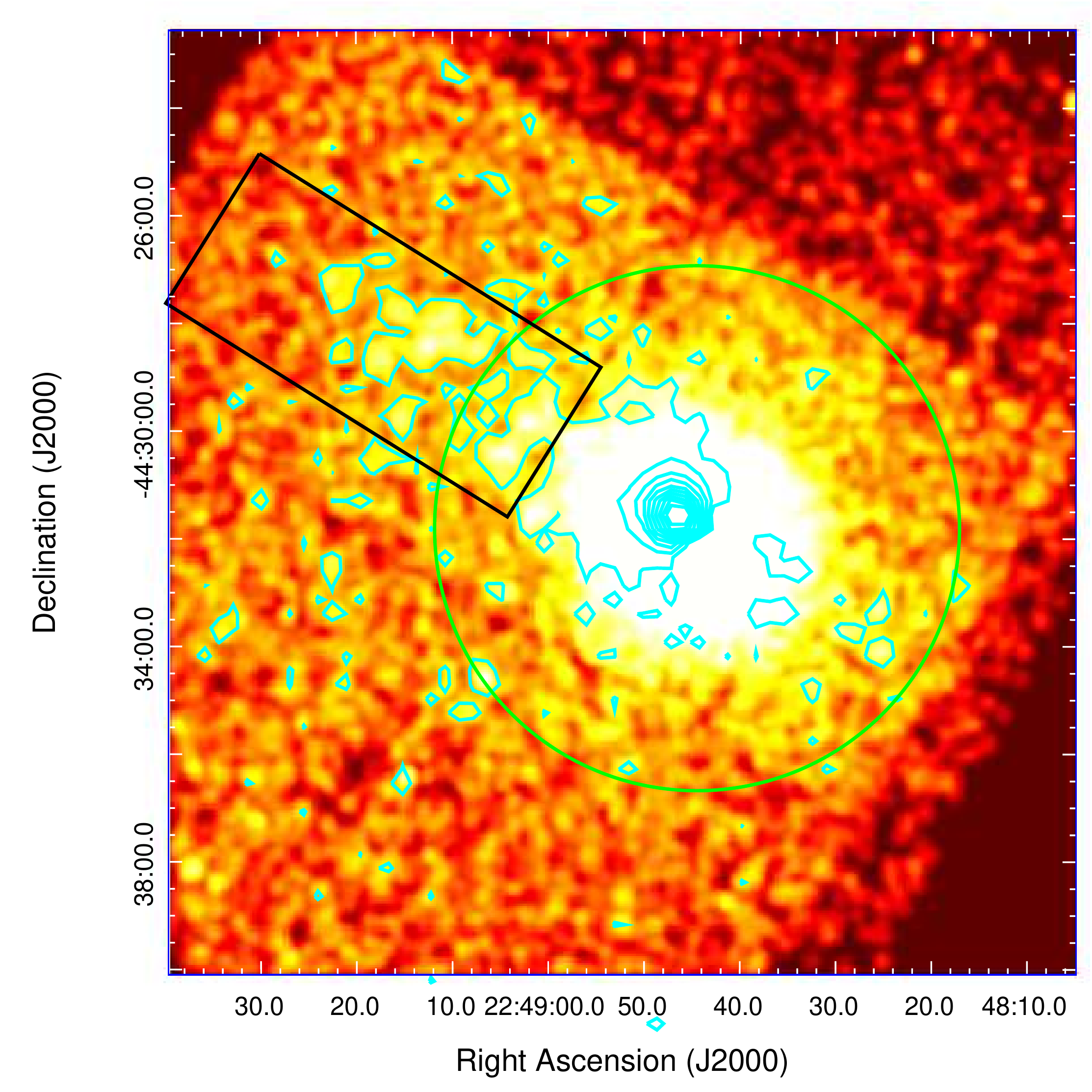} \\
\end{tabular}
\caption{Left: $\sigma$ = 4 arcsec Gaussian smoothed, \textit{Chandra} unsharp masking image (in soft energy band 0.5-1.2 keV) with respect to the best-fitting 2D elliptical double beta model. Right: \textit{Chandra} X-ray surface brightness map (in the range 0.5-2.0 keV) overlaid with contours of the \textit{Chandra} unsharp masking image (green). Box region (black) was used for spectral analysis in Section \ref{SB_excess}, yielding a cooler temperature than that of the overall cluster ICM. The green circle in both images represents the R$_{500}$ of the cluster.}
\label{fig:beta2d}
\end{figure*}

To emphasize the azimuthal asymmetries and any hidden features in the original image not seen because of the overall surface brightness gradient, we created an unsharp masking image of the cluster's surface brightness map in the soft energy band of 0.5-1.2 keV with respect to the best fitting 2D elliptical double-beta model. 
Following \citet{Ichinohe2015}, we used \textit{SHERPA} for fitting the sum of two 2D elliptical models: beta2d + beta2d. 
The center position (\textit{xpos, ypos}), ellipticities (\textit{ellip}), and the angles of the major axis (\textit{theta}) were linked between the two beta models. 
The best fit ellipticity (beta2d + beta2d model) of the cluster was found to be $0.24 \pm 0.01$, and the angle of the major axis with respect to the north is 58 degree anti-clockwise.
After subtracting the final best-fitted model from the data, the residual image is shown in Figure \ref{fig:beta2d}.

%%%%%%%%%%%%%%%%%%%%%%%%%%%% Figure X-ray residual %%%%%%%%%%%%%%%%%%%%%%%%%

\subsection{325 MHz GMRT Radio Observations}
\label{sec:radio} 
The observations of the GMRT archival data of the AS1063 cluster used in our analysis were carried out at 325 MHz on 30 July 2016 (30\_085; PI S. Hamer), 25 Feb, and 26 Aug 2017 (31\_037; PI R. J. van Weeren) for a total of about 18 hours of on-source observing time.
The data were acquired with 33.3 MHz of bandwidth divided into 512 and 256 channels, respectively.

The data analysis was performed using {\tiny SPAM} \citep{Intema2009A&A...501.1185I,Intema2017A&A...598A..78I}, and a brief description of the calibration steps are discussed below.
The calibration of the data was done in two steps.
First, the different observations were pre-calibrated separately.
Then, these data were processed together in the main pipeline. 
3C48 and 3C147 were used as flux density calibrators, and the flux density scale was set according to \citet{Scaife2012MNRAS.423L..30S}.
The pipeline performed Radio Frequency Interference (RFI) flagging and a few rounds of self-calibration followed by direction-dependent calibration on the bright sources.
Finally, a wide-field image was produced along with the calibrated visibility data. 
The output calibrated data were used for further imaging in CASA, and the final image is presented in Figure \ref{fig:Radio_maps} with Briggs robust parameter = 0 \citep{Briggs1995} and smoothed with a Gaussian beam of $23\arcsec \times 23\arcsec$, position angle = $0^{\circ}$.

\section{Results} \label{result}
\subsection{Thermodynamic maps}
\begin{figure*}
\begin{tabular}{cc}
\centering
\includegraphics[width=\columnwidth]{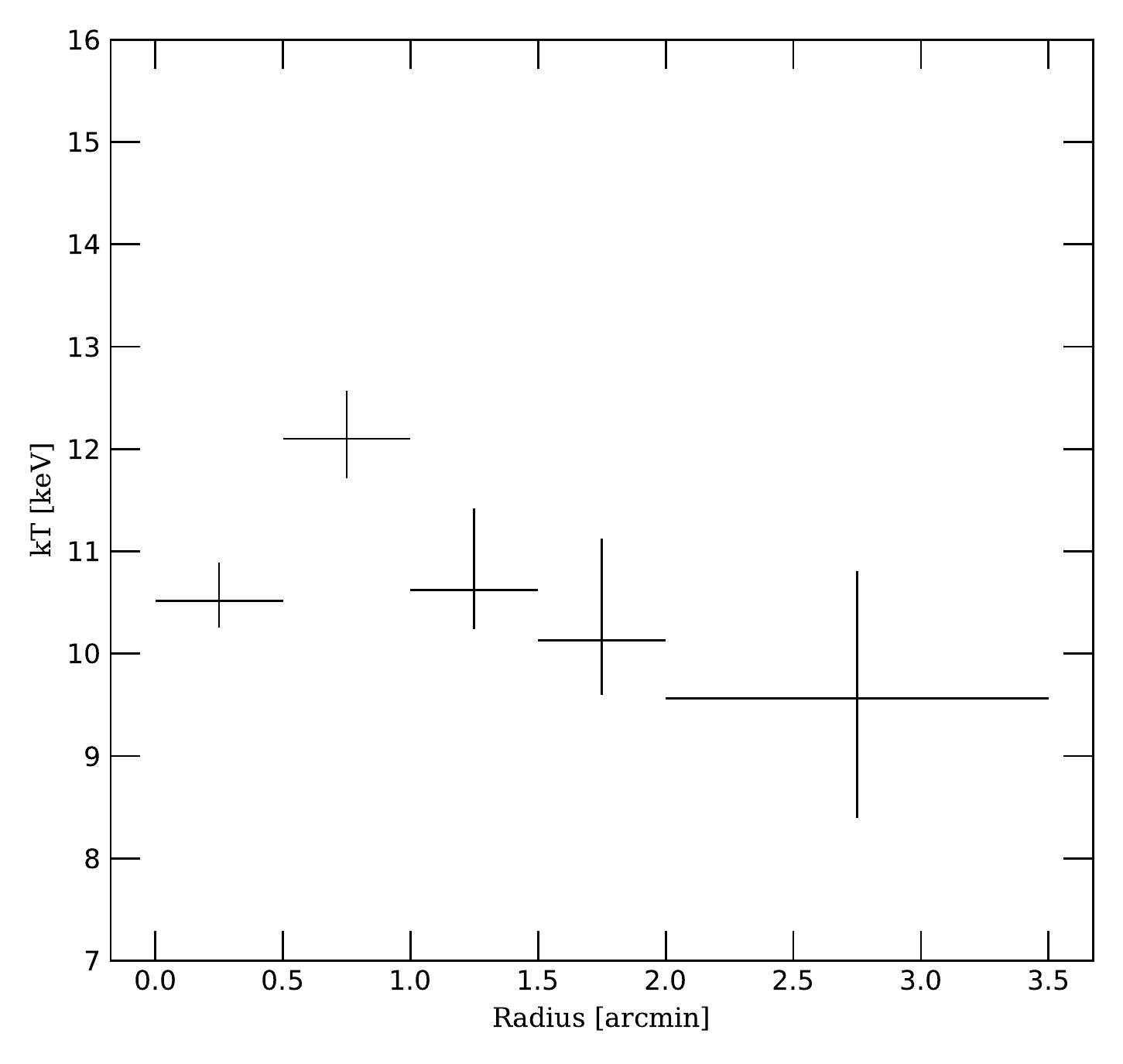} &
\includegraphics[width=\columnwidth]{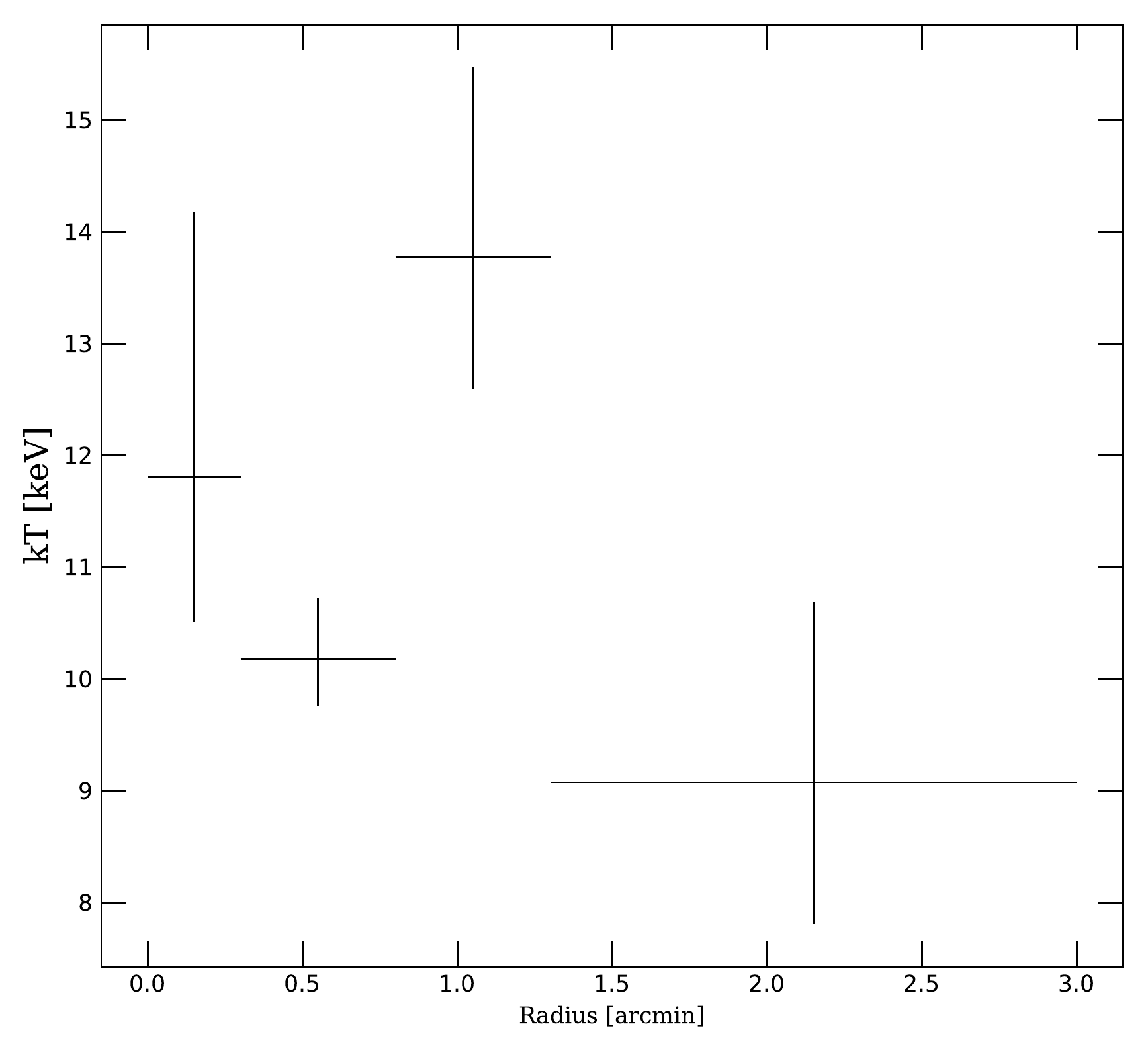} \\
\end{tabular}
\caption{Left: Average radial temperature profile of AS1063 from X-ray spectroscopy with \textit{Chandra} data in the energy range of 0.7 to 8.0 keV. Spectra were taken from annular regions and fitted using XSPEC in the same way as for the temperature map of the left panel in Figure ~\ref{fig:ACB_Tmap} (see Section \ref{sec:TMAP}).  Right: The X-ray temperature profile over the wedge region shown in the left panel of Figure \ref{fig:ACB_Tmap}. This profile was used to investigate the temperature fluctuations quantitatively as described in Section \ref{sec:Temp_dis}.}
\label{fig:T_profile}
\end{figure*}

Figure \ref{fig:X-Ray-SB} shows the projected \textit{Chandra} X-ray surface brightness map of AS1063,
which also represents the density of the cluster as the plasma density is proportional to the square-root of the X-ray surface brightness \citep{Datta2014,Schenck2014}.
In Figure \ref{fig:ACB_Tmap}, we show a projected ACB temperature map that presents a disturbed morphology.
In the temperature map, we found some hot regions as well as some comparatively lower temperature regions. 
We determined the average temperatures of different regions A, B, C, and D in Figure \ref{fig:ACB_Tmap} by extracting a separate spectrum including counts from the entire regions inside each region.
The center of the cluster is comparatively lower in temperature (although still hot compared to many other clusters); the average temperature of the central region C is found to be $10.5 \pm 1.4$ keV, and for the other regions A = $15.4 \pm 2.38$ keV, B = $14.7 \pm 2.4$ keV, and D = $12.7 \pm 2.0$ keV (Figure \ref{fig:ACB_Tmap}, left panel).
As we choose only high-temperature regions, temperatures from regions A, B, and D agree within 1$\sigma$ level. 
Temperatures from these regions differ from the comparatively lower temperature region C. To show the significance of these temperature structures, we made a dedicated temperature profile over the wedge region within one of the hot regions in Figure \ref{fig:T_profile}.
A bow-like hot region A (in Figure \ref{fig:ACB_Tmap}, left panel) is found southwest of the center, which is also aligned perpendicularly to the merger axis.
The mean temperature within $R_{500}$ ($\sim$ 1.45 Mpc, \citealt{Birzan2017MNRAS.471.1766B}) is found to be $11.68 \pm 0.56$ keV.  It should be noted that this value is consistent with a complimentary study of \textit{XMM-Newton} data resulting in a value of $T_{R500} = 11.46^{+0.28}_{-0.63}$ keV \citep{Bulbul2019ApJ...871...50B}.
The ICM temperature throughout the cluster is $\sim 11$ keV (Figure \ref{fig:T_profile}), making it one of the hottest known galaxy clusters.

%%%%%%%%%%%%%%%%%%%%%%%%%%%% Figure radial temp Profile %%%%%%%%%%%%%%%%%%%%%%

\subsection{Residual map}    \label{excess}
In Figure \ref{fig:beta2d}, we show the residual image of the X-ray surface brightness map of AS1063 in the soft energy band of 0.5-1.2 keV.
It reveals an extended excess of emission towards the northeast of the cluster, which is extended up-to $\sim$2.7 Mpc from the cluster center.
This excess brightness is also present in the medium energy band (1.2-2.0 keV), but not in the hard energy band (2.0-7.0 keV) of the \textit{Chandra} X-ray surface brightness image. 
We perhaps detect brightness excess only in the  $\leq 2$ keV band which maybe because the thermal plasma dominates at $\leq 2$ keV \citep{McDonald2019ApJ...885...63M}.
 
\subsection{Radio Halo} \label{Halo}
\begin{figure*}
\centering
\begin{tabular}{lccr}
\includegraphics[width=\columnwidth]{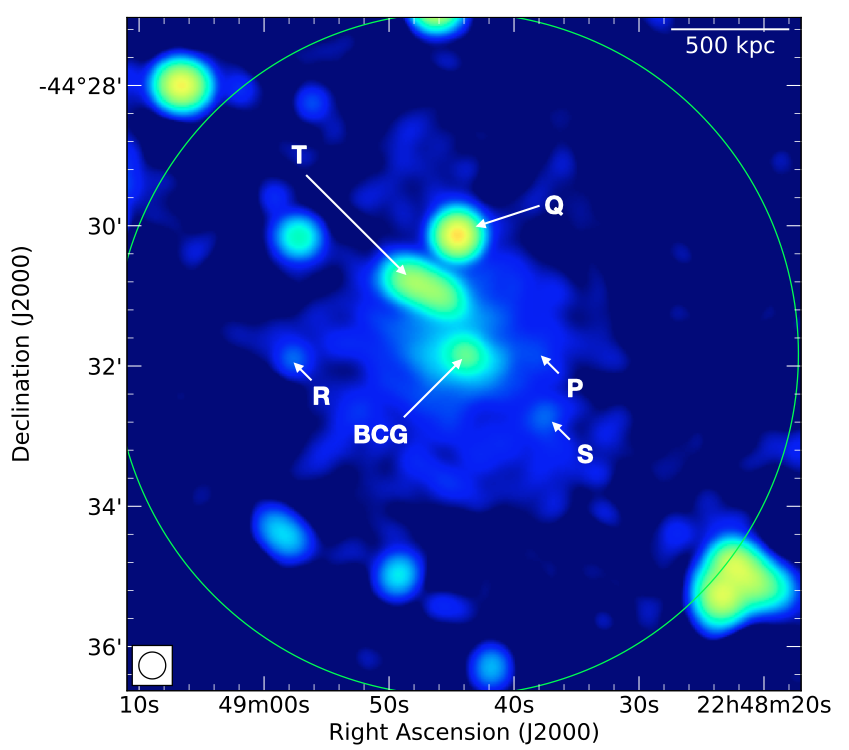} &
\includegraphics[width=\columnwidth]{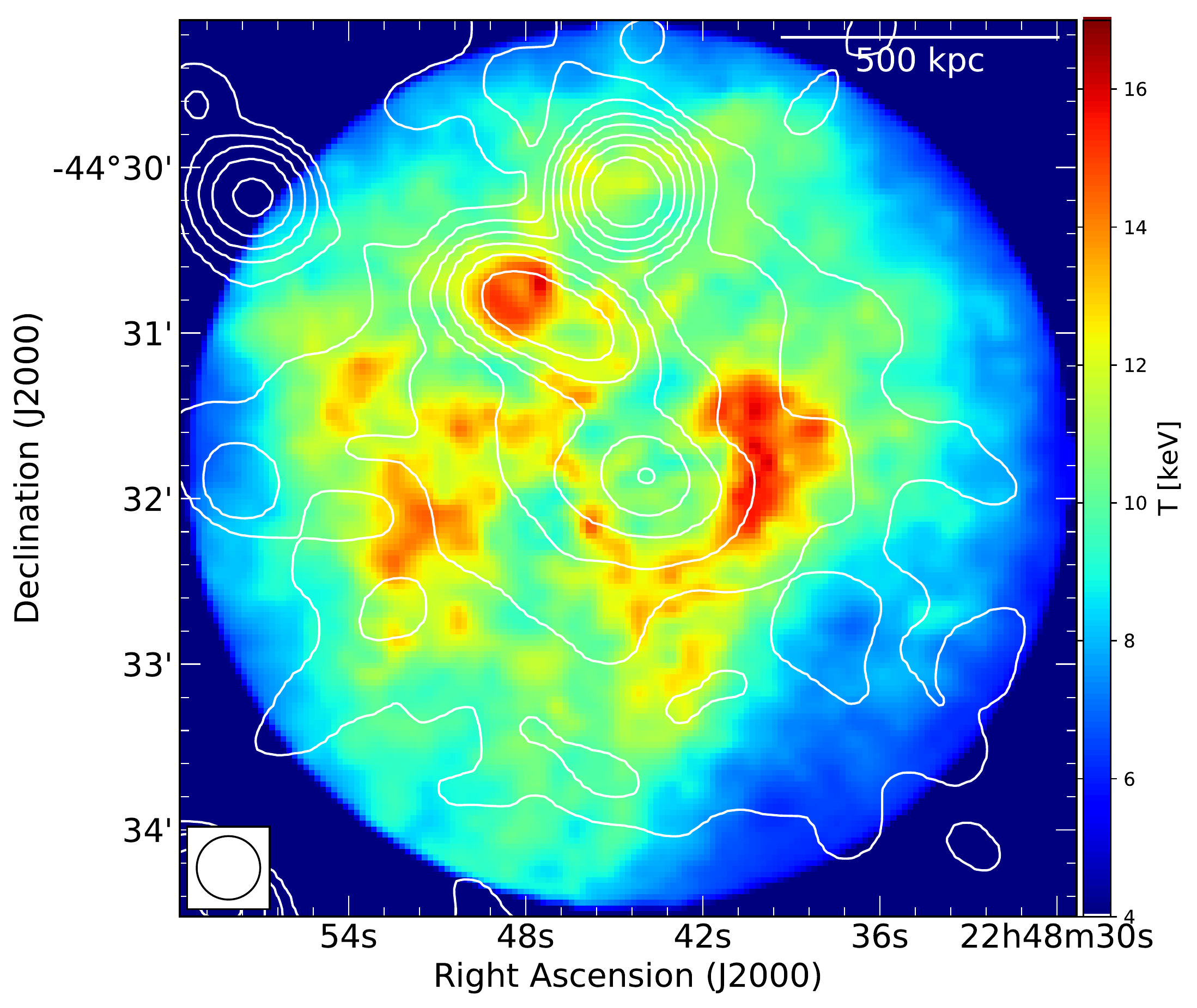} \\
\includegraphics[width=\columnwidth]{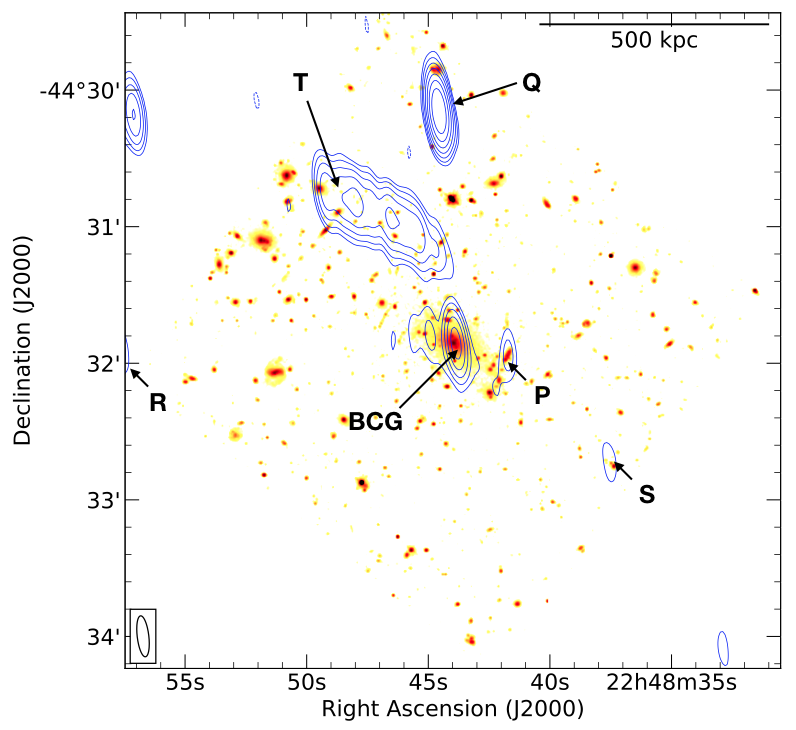} &
\includegraphics[width=\columnwidth]{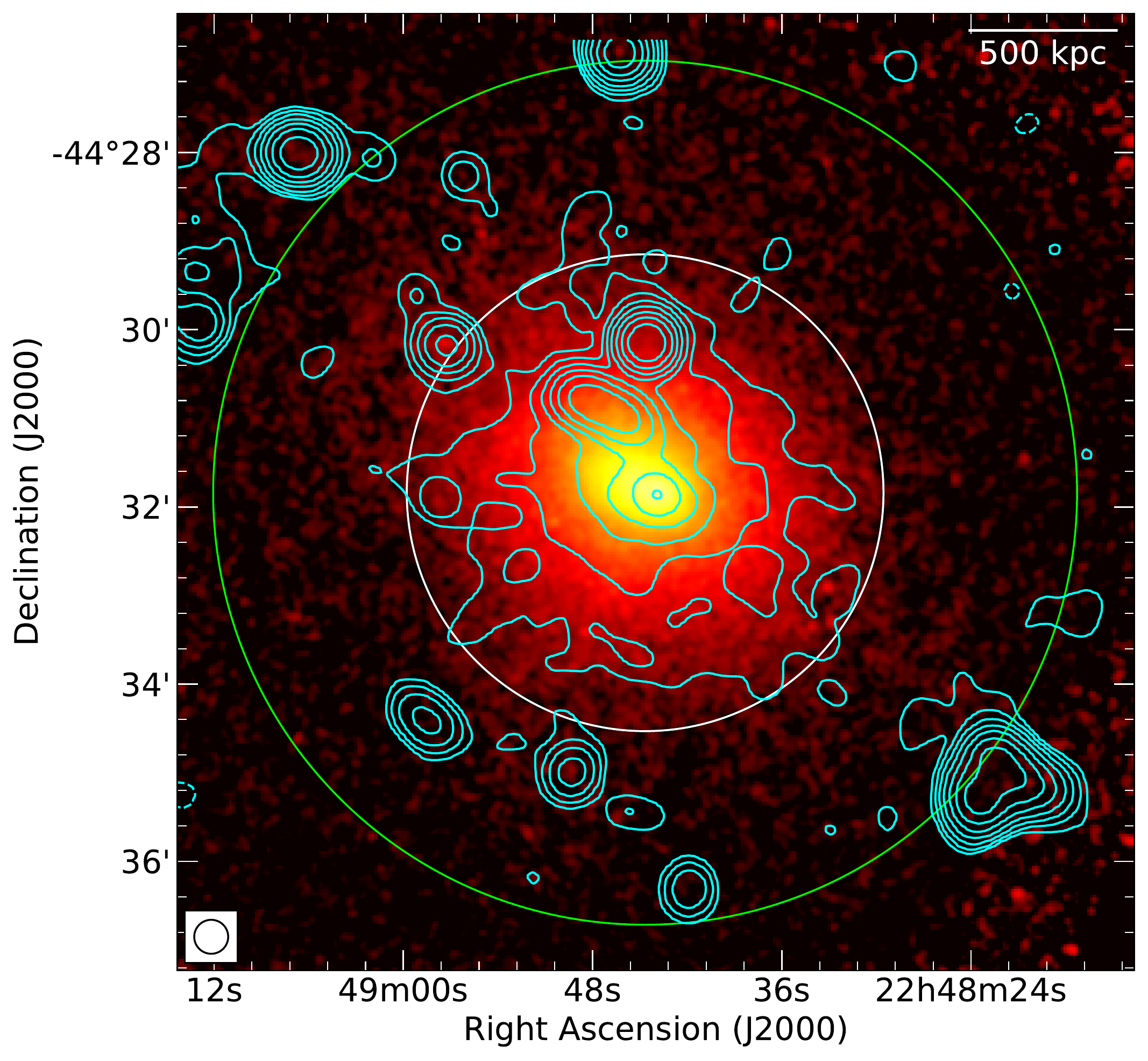} \\
\end{tabular}
\caption{Top Left: GMRT 325 MHz radio map of AS1063. Here, P, Q, R, S, and T are compact radio sources, the total flux densities of these sources are listed in Table \ref{tab:psrc}. The green circle represents R$_{500}$ of the cluster. Top Right: \textit{Chandra} ACB temperature map of AS1063 overlaid with GMRT 325 MHz radio contours in white. The contour levels are $[-1,1,2,4,8,16,32,64]\times 3\sigma_{rms}$ where $\sigma_{rms}$= 100 $\mu$Jy beam$^{-1}$.  Bottom Left: 606 nm HST optical image overlaid with radio contours in blue of the high-resolution (with 1k$\lambda$ inner \textit{uv}-cut) 325 MHz GMRT image (beam = $18.1\arcsec \times 5.1 \arcsec,\ pa = 6.8 \arcsec$). Bottom Right: X-ray surface brightness map overlaid with the same radio contours, in cyan, as those in white in the top, right panel. The green circle represents R$_{500}$, and the inner circle (cyan) represents the 800 kpc region, up to where the ACB temperature map was produced.}
\label{fig:Radio_maps}
\end{figure*}

Our analysis of 325 MHz GMRT radio data from three different observations gives a high-fidelity image of AS1063 (Figure \ref{fig:Radio_maps}).
The full resolution of the GMRT radio image at 325 MHz is about $10\arcsec$. However, to recover the diffuse radio emissions, we created a low-resolution image with a $23\arcsec \times 23\arcsec$ beam (see the top, left panel of Figure \ref{fig:Radio_maps}).
There was no significant increase in diffuse radio emission beyond the $23\arcsec \times 23\arcsec$ beam-size. 
This low-resolution image reveals large scale diffuse radio emission, covering most of the cluster region visible in X-ray. 
The bottom, right panel of Figure \ref{fig:Radio_maps} shows that the peak of the cluster's X-ray emission coincides with the peak of the 325 MHz diffuse radio emission.
At $23\arcsec$ resolution, we find that the total extent of diffuse radio emission is about $\sim$1.2 Mpc within a 3$\sigma_{rms}$ contour, where $\sigma_{rms}$ = 100 $\mu$Jy beam$^{-1}$.

We classify this radio emission as RH due to its size and location in the cluster. Some compact radio sources are also present within the RH region, which are labeled as BCG, P, Q, R, S, and T in  the top, left panel of Figure \ref{fig:Radio_maps}.
BCG is the brightest galaxy in a galaxy cluster, which is generally expected to be close to the spatial and kinematical center of the cluster, and T is possibly a Head-Tail galaxy \citep{Duchesne2017arXiv170703517D,Xie_2020A&A...636A...3X}.
To estimate the integrated flux density of the RH, the contribution from compact radio sources needs to be removed.

To subtract 
the embedded compact sources and provide a reliable measurement of flux density of the RH, we created a high-resolution image applying 1k$\lambda$ inner \textit{uv}-cut. Here, we have followed a similar procedure as described in \citet{Raja_2020ApJ...889..128R}.
1k$\lambda$ corresponds to a linear size of about 1Mpc at the cluster redshift of z = 0.351.
We applied PyBDSF \citep{Mohan2015ascl.soft02007M} to the high-resolution image to measure the integrated flux densities of the embedded compact radio sources.
The integrated flux densities of the sources, labeled as BCG, P, Q, R, S, and T, located in the RH region (see the top, left panel of Figure \ref{fig:Radio_maps}), are reported in Table \ref{tab:psrc}.
The flux density of the RH measured within 3$\sigma$ contour, and after subtracting the integrated flux densities of the compact radio sources mentioned above, is found to be $\mathrm{S_{325\ MHz}}$ = $62.0 \pm 6.28$ mJy.

The error in the total flux density is calculated as $\sqrt{(\sigma_{cal} \times S_{\nu})^2 + (\sigma_{rms} \times \sqrt{N})^2}$, where $\sigma_{cal}$ is the calibration error, $\sigma_{rms}$ is local \textit{rms} background noise and $N$ is the total number of beams within the 3$\sigma_{rms}$ contour.
We adopt a calibration error of 10\% for GMRT observations. 
We verified that the flux densities of the compact radio sources for different \textit{uv}-cuts (1.0-7.0 k$\lambda$; see Table \ref{tab:psrc}) are consistent within the error bars except for the source T, where a considerable amount of diffuse emission is present.
To exclude the maximum contribution from the source T,
we chose the image with 1 k$\lambda$ inner \textit{uv}-cut for subtraction; because, in this image, the source T contains more diffuse emission.

%point source table
\begin{table*}
	\centering
    \caption{The integrated flux densities of the compact radio sources corresponding to different inner \textit{uv}-cuts are listed below. These sources are shown in Figure \ref{fig:Radio_maps}.
    At the cluster redshift of 0.351, 1k$\lambda$ corresponds to $\sim$206 arcsec of angular scale, which is about 1Mpc.
    }
    \label{tab:psrc}
	\begin{tabular}{lccccr}
	\hline
    Compact radio &  & 7k$\lambda$ \textit{uv}-cut &	 5k$\lambda$ \textit{uv}-cut &	 3k$\lambda$ \textit{uv}-cut &	 1k$\lambda$ \textit{uv}-cut \\
    sources & Identifier & flux density & flux density & flux density & flux density \\
     & & (mJy) & (mJy) & (mJy) & (mJy) \\
    \hline
    BCG & 2MASX J22484405-4431507 & 	5.18 $\pm$ 0.15 	&	6.48 $\pm$ 0.13   & 	6.71 $\pm$ 0.12   &     6.61 $\pm$ 0.11 \\
    P   & rxj2246\_18112 &	0.84 $\pm$ 0.16 	&	0.90 $\pm$ 0.13   & 	0.77 $\pm$ 0.11   & 	0.54 $\pm$ 0.10 \\
    T  & rxj2248\_18479 &	5.47 $\pm$ 0.33 	&	25.72 $\pm$ 0.41  & 	33.89 $\pm$ 0.38  &	    36.48 $\pm$ 0.40 \\
    Q   & [GVR2012] 1472 &	39.63 $\pm$ 0.24	&	41.24 $\pm$ 0.19  & 	42.62 $\pm$ 0.18  &	    42.49 $\pm$ 0.18 \\
    R   & [GVR2012] 119 &	0.92 $\pm$ 0.13     & 	0.91 $\pm$ 0.12   &	    0.92 $\pm$ 0.11   & 	0.93 $\pm$ 0.11 \\
    S   & [GVR2012] 844 & &   &                                       	0.55 $\pm$ 0.12   &	    0.58 $\pm$ 0.12 \\

        \hline
	\end{tabular}
\end{table*}

%%%%%%%%%%%%%%%%%%%%%%%%%%%% Figure Radio map %%%%%%%%%%%%%%%%%%%%%%%%%

\section{Discussion} \label{sec:discussion}
\subsection{Dynamical state of the cluster}
The presence of diffuse radio emission in any cluster directly depends upon the dynamical state of that cluster. Almost all RHs and radio relics are detected in merging clusters, and mini-halos are found in relaxed cool core clusters \citep{Wen2013MNRAS.436..275W}.
To investigate the dynamical state of AS1063, we examined various observables using \textit{Chandra} X-ray data as described below.

\subsubsection{Dynamical Indicators}    \label{sec:DS}
To investigate the dynamical state of AS1063, we looked into  some well known morphology classification indicators. All parameters are listed in Table \ref{tab:parameters}.
Cluster dynamical state classifies whether the cluster is merging (disturbed) or relaxed (undisturbed).
Photon asymmetry ($\mathrm{A_{phot}}$; \citealt{Nurgaliev2013ApJ...779..112N}) and centroid shift (\textit{w}; \citealt{Mohr1993ApJ...413..492M})  are used to estimate disturbance in clusters (e.g., \citealt{Poole2006MNRAS.373..881P,OHara2006ApJ...639...64O,Ventimiglia2008ApJ...685..118V,Maughan2008ApJS..174..117M,Bohringer2010,cassano2013ApJ...777..141C}). 
For AS1063, \textit{w} = $0.006^{+0.001}_{-0.000}$ \citep{McDonald2013ApJ...774...23M}, which falls in the relaxed category according to the criteria classifying clusters as relaxed if $w < 0.012$ and merging otherwise \citep{Cassano2010A&A...517A..10C}, and $\mathrm{A_{phot}}$ = 0.21 which falls in the moderate asymmetry range (0.15 - 0.6) corresponding to a moderately disturbed cluster \citep{Nurgaliev2017ApJ...841....5N}. Note that, $\mathrm{A_{phot}}$ has more statistical power in resolving substructure and is able to produce more consistent results independently of the quality of the data than \textit{w} \citep{Nurgaliev2017ApJ...841....5N}.

Also, we used the surface brightness concentration parameter $\mathrm{c_{SB}}$ \citep{Santos2008}, the central cooling time $\mathrm{t_{cool,0}}$ and the entropy $\mathrm{K_0}$ \citep{Cavagnolo2009ApJS..182...12C} to investigate whether the cluster has a cool core (CC) or not.
We calculated $\mathrm{c_{SB}} = 0.23 \pm 0.01$, which suggests that AS1063 is a CC cluster\footnote{\citet{Santos2008} defined three categories of clusters: (a) non CC ($\mathrm{c_{SB}} < 0.075$), (b) moderate CC ($0.075 < \mathrm{c_{SB}} < 0.155$), and (c) pronounced (strong) CC ($\mathrm{c_{SB}} > 0.155$).}.
\citet{Hudson2010A&A...513A..37H} classified\footnote{\citet{Hudson2010A&A...513A..37H} divided clusters into three types: strong CC ($\mathrm{t_{cool,0}} < 1 \mathrm{h^{-1/2}}$ Gyr \& $\mathrm{K_0 \leq 30 h^{-1/3}}$ keV cm$^2 $), weak CC ($\mathrm{t_{cool,0}}$ between 1-7.7 $\mathrm{h^{1/2}}$ Gyr \& $\mathrm{K_0 \geq 30 h^{-1/3} keV cm^2 }$) and non-CC clusters ($\mathrm{t_{cool,0}} > 7.7 \mathrm{h^{-1/2}}$ Gyr \& $\mathrm{K_0 > 110 h^{-1/3}}$ keV cm$^2 $).} clusters using $\mathrm{t_{cool,0}}$.
For AS1063, $\mathrm{t_{cool,0}}$ = 1.79  $\mathrm{h^{-1/2}}$ points to being a weak CC\footnote{``Weak cool-core clusters have enhanced central entropies and temperature profiles that are flat or decrease slightly towards the center'' - \citet{Hudson2010A&A...513A..37H}.} cluster whereas $\mathrm{K_0}$ = $128.0^{+9.5}_{-9.4}$ keV cm$^2$ [or $113.65.0^{+8.435}_{-8.346}\mathrm{h^{-1/3}}$ keV cm$^2 $] situates it just above the weak CC to non-CC boundary. 
On the other hand, a high mass deposition rate of $\mathrm{dM/dt_{7.7}}$ = $652.4^{+48.2}_{-42.0}\ \mathrm{M_{\odot}\ yr^{-1}}$ also indicates the presence of a cool core in the cluster. 
\citet{Lovisari2017ApJ...846...51L} report A1063 as relaxed based on morphological parameters, e.g., high concentration parameter, low power ratio, and low centroid shift of the X-ray emission. 
Therefore, this shows that such dynamical indicators can be misleading if the merger happens to be along the line of sight \citep{Xie_2020A&A...636A...3X}.
In the next section, we present instead a more insightful view of the dynamical state of the cluster using a high-quality temperature map.

\subsubsection{Temperature map} \label{sec:Temp_dis}
The \textit{Chandra} X-ray surface brightness map in Figure \ref{fig:X-Ray-SB} shows a smooth distribution except for some elongation in the northeast-southwest direction.
However, in the ACB temperature map of the left panel in Figure \ref{fig:ACB_Tmap}, some hot regions appear (A, B, and D in Figure \ref{fig:ACB_Tmap}) as well as comparatively lower temperature regions.
It is well known that the sensitivity of the Chandra X-ray Observatory degrades significantly above 10 keV \citep{Datta2014}. The resultant temperatures in the ACB map for AS1063 are high and in some regions they correlate with high errors as shown in the right panel of Figure \ref{fig:ACB_Tmap}.
To verify these fluctuations in the ACB temperature map, we create another temperature map using an alternate technique, the Weighted Voronoi Tessellation (WVT) method, as described in  Section \ref{wvt} and shown in Figure \ref{fig:WVT}. 
Note that the high temperature regions in the ACB map of the left panel in Figure \ref{fig:ACB_Tmap} overlap with the high temperature regions in the WVT map of the top panel of Figure \ref{fig:WVT}. Moreover, for quantitative comparison, we plot the best-fitted temperatures from both temperature maps, which are in good agreement with a correlation coefficient of 0.85 (see Figure \ref{fig:temp_correlation}). This reinforces the significance of the higher temperature values and the overall ICM temperature fluctuations within the cluster region of AS1063.

To study temperature variation quantitatively, we make an average radial (azimuthally symmetric) temperature profile (see the left panel of Figure \ref{fig:T_profile}). This profile shows a significant rise in temperature from the centre of the cluster to a distance of $\sim 0.8$~arcmin and then gradual decrease in temperature near the edge of the cluster. This rise in temperature is more evident in the temperature profile along the wedge drawn across region A in the left panel of Figure~\ref{fig:ACB_Tmap}. This profile is shown in the right panel of Figure \ref{fig:T_profile}.
The disturbed morphology in the temperature map is a signature of being a merging cluster \citep{Burns2016AAS...22811002B,Hallman2018}.
The mean temperature of the cluster is very high, i.e., $T_{500} = 11.68\pm 0.56$ keV, which might also be due to significant merging activity in the cluster.

\subsubsection{X-ray brightness excess} \label{SB_excess}
Using simulations, \citet{Gomez2012AJ....144...79G} showed that AS1063 is undergoing a Bullet-like recent merger event (1:4 mass ratio merger) close to the plane of the sky.
In the optical image (Figure \ref{fig:Radio_maps}), the central BCG coincides with the X-ray peak, which is common in cool core clusters. But, there is no sign of a cool core in the temperature map (see Figure \ref{fig:ACB_Tmap}) or in the temperature profile (see Figure \ref{fig:T_profile}). 
In our analysis, the axis of the excess in X-ray brightness map (Figure~\ref{fig:beta2d}) is found to be 58$^{\circ}$ anticlockwise from the north, which is consistent with the previously reported merging axis by \citet{Gomez2012AJ....144...79G}.

The direction of this excess brightness is the same as that of the merging axis suggested by \citet{Gomez2012AJ....144...79G}, where the merging axis was predicted using N-body simulations and velocity distribution in the cluster. Therefore, the brightness excess found in our X-ray residual map confirms the previously predicted merging scenario by providing
observational evidence of the merging scenario in AS1063.
We fitted spectra within a rectangular region of $1 \times 2.48$ Mpc$^2$ area (see Figure \ref{fig:beta2d}), situated at about 1.0 Mpc (where, $R_{500}$ = 1.45 Mpc) distance from the center of this cluster along the ``brightness excess'' region.
We found a temperature of $1.71^{+0.84}_{-0.32}$ keV (using the APEC thermal model in XSPEC) in the 0.7-2.0 keV energy band (X-ray emission at $> 2$ keV is much less in this brightness excess region), which is cooler than the cluster ICM temperature ($\sim 11 keV$).
As the excess region expands beyond R$_{500}$, the count rate falls drastically, which may lead to underestimating the derived temperature by a factor of $\sim 3$ \citep{Leccardi_2008A&A...486..359L}. 

In the residual map of AS1063 (see Figure \ref{fig:beta2d}), the excess region's brightness gradually increases towards the cluster center, which may indicate that we are witnessing the infall of the subcluster in the potential well of the main cluster (see Figure \ref{fig:Box_a_b}). Therefore, we assume that this brightness excess is associated with the stripped cool gas left behind by the merging subcluster.
This interpretation is supported further by our measurement of the gas temperature ($\sim$1.7 keV), which is significantly lower than the ambient ICM ($\sim$11 keV) and typical of the plasma of a galaxy group/subcluster with a mass of a few $10^{13} M_\odot$ \citep{Eckert_2014A&A...570A.119E}.
Further, as the cluster is elongated towards the excess emission, we also consider the possibility of the excess emission being part of a cosmic filament.

\begin{figure}
\centering
\includegraphics[width=\columnwidth]{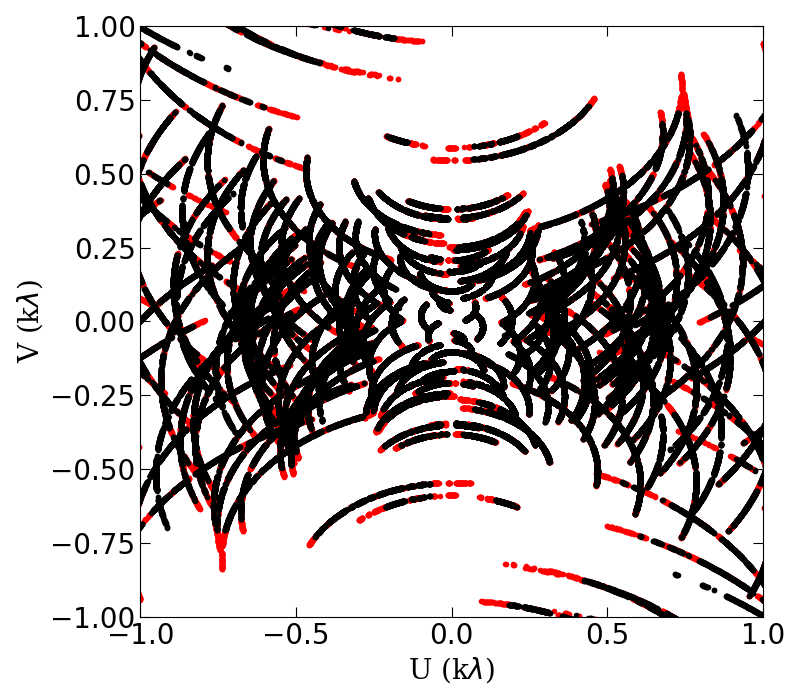}
\caption{ UV-plot of combined 325 MHz GMRT observations, where black represents data presented in \citet{Xie_2020A&A...636A...3X} (PI: van Weeren) and red represents additional data presented in our analysis (PI: Hamer).}
\label{fig:uv-plot}
\end{figure}

\subsection{Spectral Property of the Radio Halo} \label{sec:radio_spectra}
\begin{figure}
\centering
\includegraphics[width=\columnwidth]{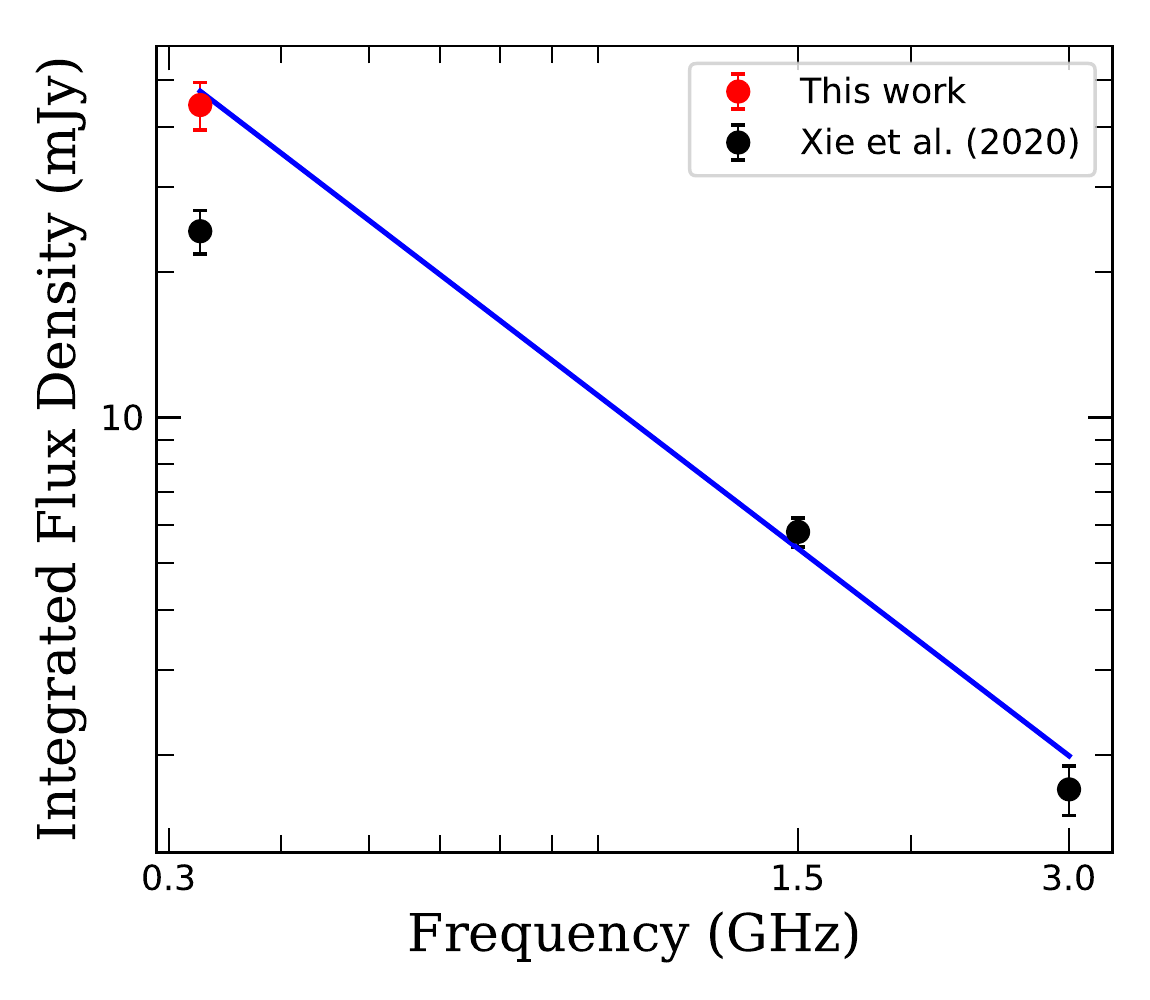}
\caption{Observed Integrated Radio Flux density of the RH in AS1063 across different radio frequencies. The measurements represented by black dots are taken from \citet{Xie_2020A&A...636A...3X}, the red dot represents the new flux density from our analysis.  The blue line is the  best  power-law  fit  to  the  data, with a spectral index of $\alpha = -1.43 \pm 0.13$.% (reduced-$\chi^2$ = 0.27).
}
\label{fig:halo_spectrum}
\end{figure}

The spectral properties of the RH are important to understand its origin
\citep{van_Weeren2019SSRv..215...16V}.
In Figure \ref{fig:halo_spectrum}, we present the integrated radio spectrum between 325 MHz to 3 GHz. The integrated flux density at 1.5 and 3.0 GHz are taken from \citet{Xie_2020A&A...636A...3X}. They also measured an integrated
325 MHz flux of $24.3 \pm 2.5$ mJy, which is considerably lower than our measured flux density ($44.34\pm 4.97$ mJy).
 The discrepancy is likely due to the data used in both studies. We have used a significantly larger dataset with improved \textit{UV} coverage, resulting in a better recovery of the diffuse radio emission (see Figure \ref{fig:uv-plot}).
To check the reliability of our reported value, we calculated the integrated flux density from images using individual observations as well as combined observations and we found consistent results.
We also tried the same imaging parameters used in \citet{Xie_2020A&A...636A...3X}, which yielded the same integrated flux density as we report here. We carefully subtracted integrated flux density contributions of the embedded compact radio sources from the total diffuse emission, as discussed in Section \ref{Halo}.
Adding our new measurement to those presented in  \citet{Xie_2020A&A...636A...3X}, the data can be fitted well with a single power law of
spectral index\footnote{Note that the spectral index may vary slightly as the region we used to extract RH flux density is not the exact region reported in \citet{Xie_2020A&A...636A...3X}. However, we did try to make the region as close as possible.} $\alpha = -1.43 \pm 0.13$ (see Figure \ref{fig:halo_spectrum}).

%%%%%%%%%%%%%%%%%%%%%%%%%%%% Halo spectrum %%%%%%%%%%%%%%%%%%%%%%%%%%%%%%%%%

%%%%%%%%%%%%%%%%%%%%%%%%%%%% P1.4-Lx %%%%%%%%%%%%%%%%%%%%%%%%%%%%%%%%%
\begin{figure}
\centering
\includegraphics[width=\columnwidth]{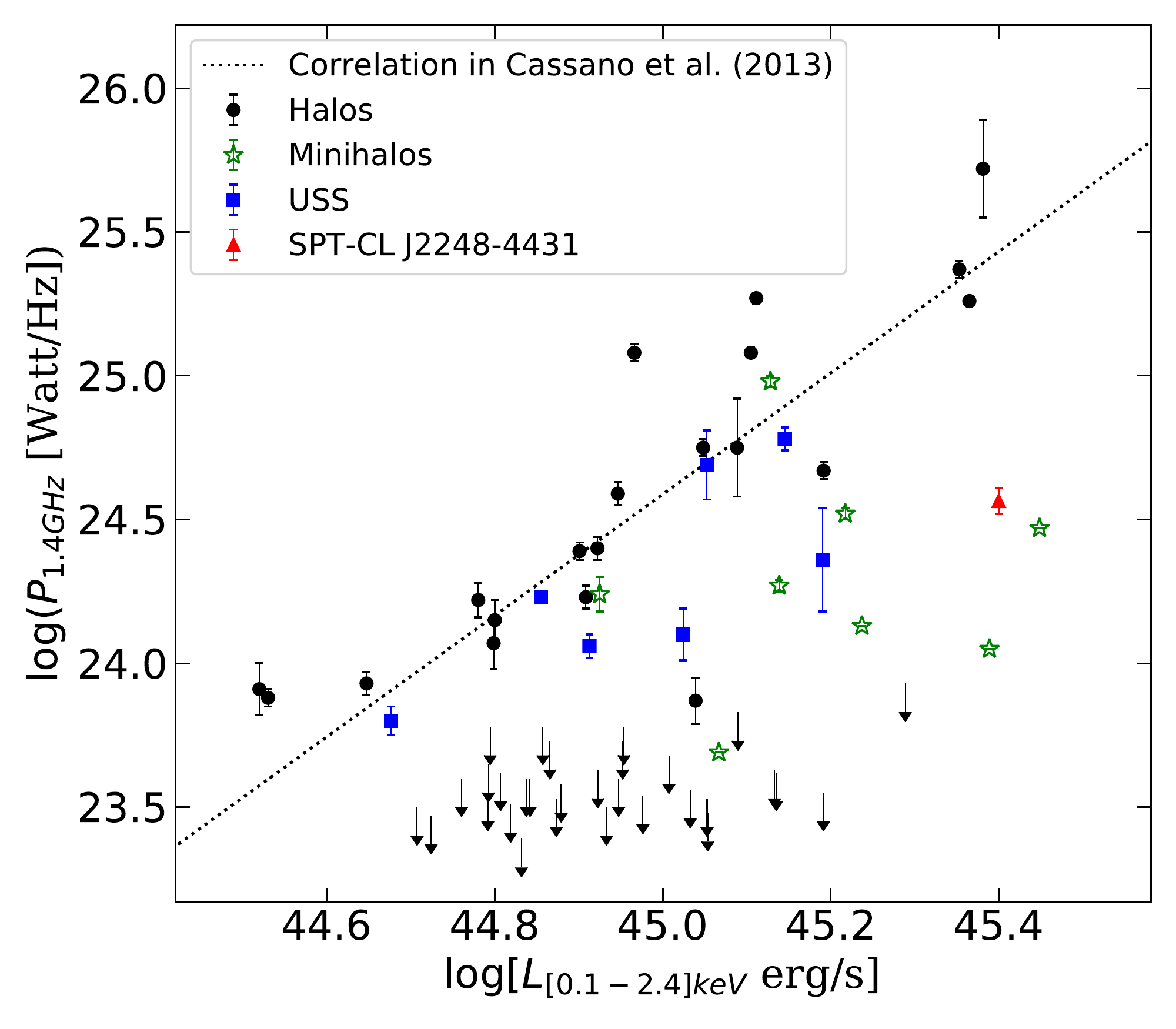}
\caption{$L_X-P_{1.4\ \mathrm{GHz}}$ scaling relation data (taken from \citealt{cassano2013ApJ...777..141C}) updated with the RH in AS1063 (red triangle). The dotted line is the correlation reported in \citet{cassano2013ApJ...777..141C}.}
\label{fig:P14_LX}
\end{figure}

\subsection{Origin of the RH}
As mentioned in the introduction (Section \ref{intro}), the origin of RHs is unclear with two proposed models: turbulence re-acceleration and hadronic.
Turbulent re-acceleration is directly related to ICM turbulence created by a cluster merger.
AS1063 shows multiple signatures of an ongoing merger in the \textit{Chandra} X-ray analysis, which includes an elongated morphology in the surface brightness map, a patchy ACB temperature map, X-ray brightness excess along the merger axis, and high average ICM temperature.
Dynamical state indicators also suggest that AS1063 should be a moderately disturbed non-CC cluster.
So, the RH present in AS1063 may support the turbulent re-acceleration model.

Previous studies demonstrated an empirical relation between $P_{1.4\ \mathrm{GHz}}$ and $L_\mathrm{X}$ \citep{cassano2013ApJ...777..141C}. 
 We extrapolated the RH flux density with an spectral index of $\alpha=-1.43$ (see Section \ref{sec:radio_spectra}) to calculate $P_{1.4\ \mathrm{GHz}}$. 
We computed a k-corrected $P_{1.4\ \mathrm{GHz}}$ using 

\begin{equation}
    \mathrm{P_{1.4\ GHz} = \frac{4\pi D^{2}_L(z)}{(1 + z)^{(\alpha + 1)}} S_{1.4\ GHz}\ [W\ Hz^{-1}]}\,,
    \label{eq:P1.4GHz_power}
\end{equation}

\noindent
where $D_L(z)$ is the luminosity distance at the redshift of the cluster.
 The k-corrected $P_{1.4\ \mathrm{GHz}}$ is found to be $3.67 \pm 0.37 \times 10^{+24}$ W/Hz.
We plotted $P_{1.4\ \mathrm{GHz}}$ against $L_X$ ($2.51^{+0.02}_{-0.03} \times 10^{45}$ ergs s$^{-1}$ in 0.1-2.4 keV energy band) in Figure \ref{fig:P14_LX}.
We found that this RH is situated in the ultra-steep spectrum RH region.
The spectral index of $\alpha = -1.43 \pm 0.13$ obtained from the fit shown in Figure \ref{fig:halo_spectrum} is steeper than the average spectral index of RHs ($\sim -1.2 \pm 0.2$; \citealt{Feretti_2012A&ARv..20...54F}).
An RH created by the turbulent re-acceleration model is expected to produce steeper spectra \citep{Cassano_2009ASPC..407..223C}. Hence, the steep radio spectrum of the RH in AS1063 indicates a preference for the re-acceleration model over the hadronic model \citep{Brunetti_2004JKAS...37..493B}.
According to the Figure~\ref{fig:P14_LX}, the RH present in AS1063 is significantly under-luminous compared to the radio power of classical RHs hosted in clusters with similar X-ray luminosity.  It is a factor of $\sim$9 below the correlation in \citet{cassano2013ApJ...777..141C} for the given $L_X$, suggesting a hadronic origin. 
As AS1063 is experiencing a recent merger \citep{Gomez2012AJ....144...79G}, the RH may have resulted from turbulent re-acceleration within the cluster. Considering an evolutionary effect, this RH may be moving up towards the $L_{X}-P_{1.4\ \mathrm{GHz}}$ correlation (thus, switching on stage as it is experiences a recent merger, \citealt{Donnert2013MNRAS.429.3564D}).

\section{conclusions} \label{sec:conclude}
In this paper, we have analyzed 325 MHz GMRT and \textit{Chandra} X-ray observations of the moderately disturbed cluster AS1063.
The key results from our analyses are listed below:

\begin{itemize}
\item We presented a high-resolution \textit{Chandra} X-ray projected ACB temperature map, which shows that the ICM temperature is very high throughout the cluster (see temperature profile in the left panel of Figure \ref{fig:T_profile}).
The mean temperature of the cluster within $R_{500}$ is $11.68 \pm 0.56$ keV,  which indicates that the cluster accreted a lot of mass via the cluster merger.
This high average temperature and the disturbed structure in the temperature map (left panel in Figure \ref{fig:ACB_Tmap}) are good indicators of the ongoing merging process in the cluster.

\item We reported a very significant ``brightness excess" in the unsharped mask residual of the X-ray surface brightness map along the northeast-southwest direction, which might be caused by the stripped gas from the merging subcluster. This brightness excess represents a revealing signature of a merging event (Section \ref{sec:filament}).
The axis of this excess is found to be $\sim 58^{\circ}$ anticlockwise from the north, which is consistent with the previously predicted merging axis by \citet{Gomez2012AJ....144...79G} (see Section \ref{SB_excess}).
This is the first observational confirmation using X-ray of the merging scenario in AS1063 suggested by \citet{Gomez2012AJ....144...79G}.

\item We analysed archival GMRT radio observations of the RH in AS1063. The total integrated radio flux density of the RH is found to be  $\mathrm{S}[325MHz] = 62.0\pm6.3$ mJy. This flux density was estimated after careful subtraction of the embedded radio sources.
A single power-law fit to the data in flux density and frequency space yields a spectral index of $\alpha = -1.43 \pm 0.13$, which is steeper than the average spectral index of RHs ($\sim -1.2 \pm 0.2$; \citealt{Feretti_2012A&ARv..20...54F}).
This steep spectrum supports the turbulent re-acceleration model as the physical mechanism powering radio emission in RHs. 

\item The estimated radio power $P_{1.4\ \mathrm{GHz}}$ places the RH in the under-luminous ultra-steep spectrum RH region of the $L_X-P_{1.4\ \mathrm{GHz}}$ plane (Figure \ref{fig:P14_LX}).
The RH is found to be $\sim$9 times below the $L_X-P_{1.4\ \mathrm{GHz}}$ correlation given in \citet{cassano2013ApJ...777..141C}.
This  implies two possibilities:  
a) the RH may be generated by electrons originating from hadronic collisions (\citealt{Brunetti2011MNRAS.410..127B}), or 
b) the RH may result from turbulent re-acceleration within the ICM due to the recent merger, and may therefore be in the switching-on stage and moving towards the observed correlation (\citealt{Donnert2013MNRAS.429.3564D}).

\end{itemize}

\section*{Acknowledgements}
We would like to thank the anonymous reviewer for suggestions and comments that have helped to improve this paper.
MR and RR would like to thank Manoneeta Chakraborty, Arnab Chakraborty, Sumanjit Chakraborty, and Ruta Kale for fruitful discussions. 
MR acknowledges the support provided by the DST-Inspire fellowship program (IF160343) by DST, India. 
RR is supported through grant ECR/2017/001296 awarded to AD by DST-SERB, India. 
This work was also supported by NASA ADAP grant NNX15AE17G to JB. 
This work, via DR, was also supported by NASA (National Aeronautics and Space Administration) under award number NNA16BD14C for NASA Academic Mission Services.
We thank IIT Indore for providing computing facilities for the data analysis. 
We also thank both GMRT and VLA staff for making these radio observations possible. GMRT is run by the National Centre for Radio Astrophysics of the Tata Institute of Fundamental Research. 
The scientific results from the X-ray observations reported in this article are based on data obtained from the \textit{Chandra} X-ray Observatory (CXO). 
This research has also made use of the software provided by the \textit{Chandra} X-ray Center (CXC) in the application packages CIAO, ChIPS, and Sherpa. 
This research made use of APLpy \citep{aplpy2012,aplpy2019}, an open-source plotting package for Python.

%%%%%%%%%%%%%%%%%%%%%%%%%%%%%%%%%%%%%%%%%%%%%%%%%%



\section*{DATA AVAILABILITY}
The X-ray data underlying this article are available in \textit{Chandra} data archive at [https://cda.harvard.edu/chaser/mainEntry.do], and can be accessed with Obs-IDs 4966, 18611, 18818.
The radio data used in this article are available in GMRT archive at [https://naps.ncra.tifr.res.in/goa/data/search] with proposal ID 30\_085; PI - S. Hamer, and 31\_037; PI - R. J. van Weeren.
The optical data used here are available at [http://archive.stsci.edu/hst/], and can be accessed with JCQTA1020.

\bibliographystyle{mnras}
\bibliography{reference} % if your bibtex file is called example.bib

%\begin{thebibliography}{99}

%\end{thebibliography}

%%%%%%%%%%%%%%%%%%%%%%%%%%%%%%%%%%%%%%%%%%%%%%%%%%

%%%%%%%%%%%%%%%%% APPENDICES %%%%%%%%%%%%%%%%%%%%%

\appendix

\section{Weighted Voronoi Tessellation Method}   \label{wvt}
\begin{figure}
\centering
\begin{tabular}{c|c}
\includegraphics[width=\columnwidth]{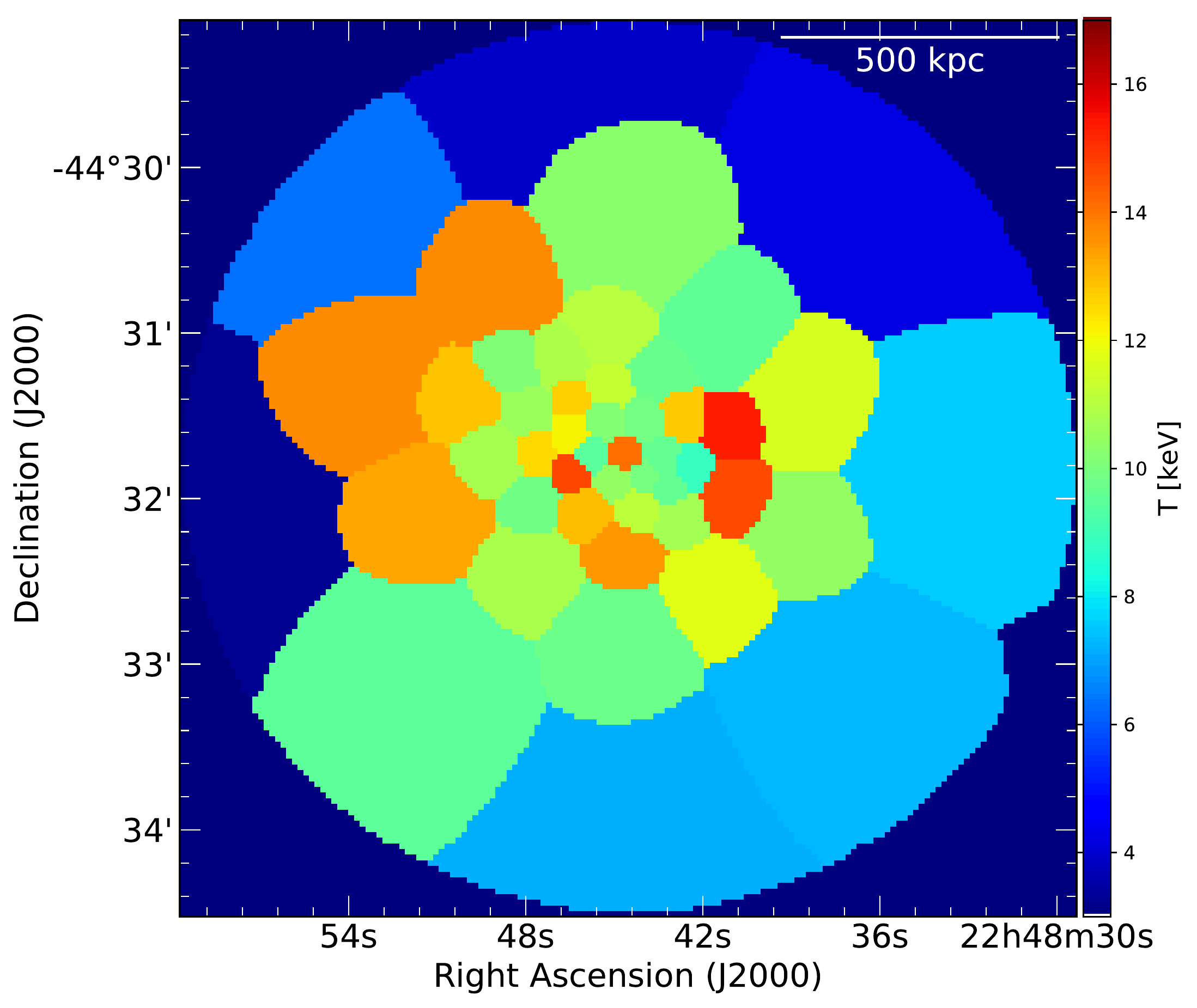} \\
\includegraphics[width=\columnwidth]{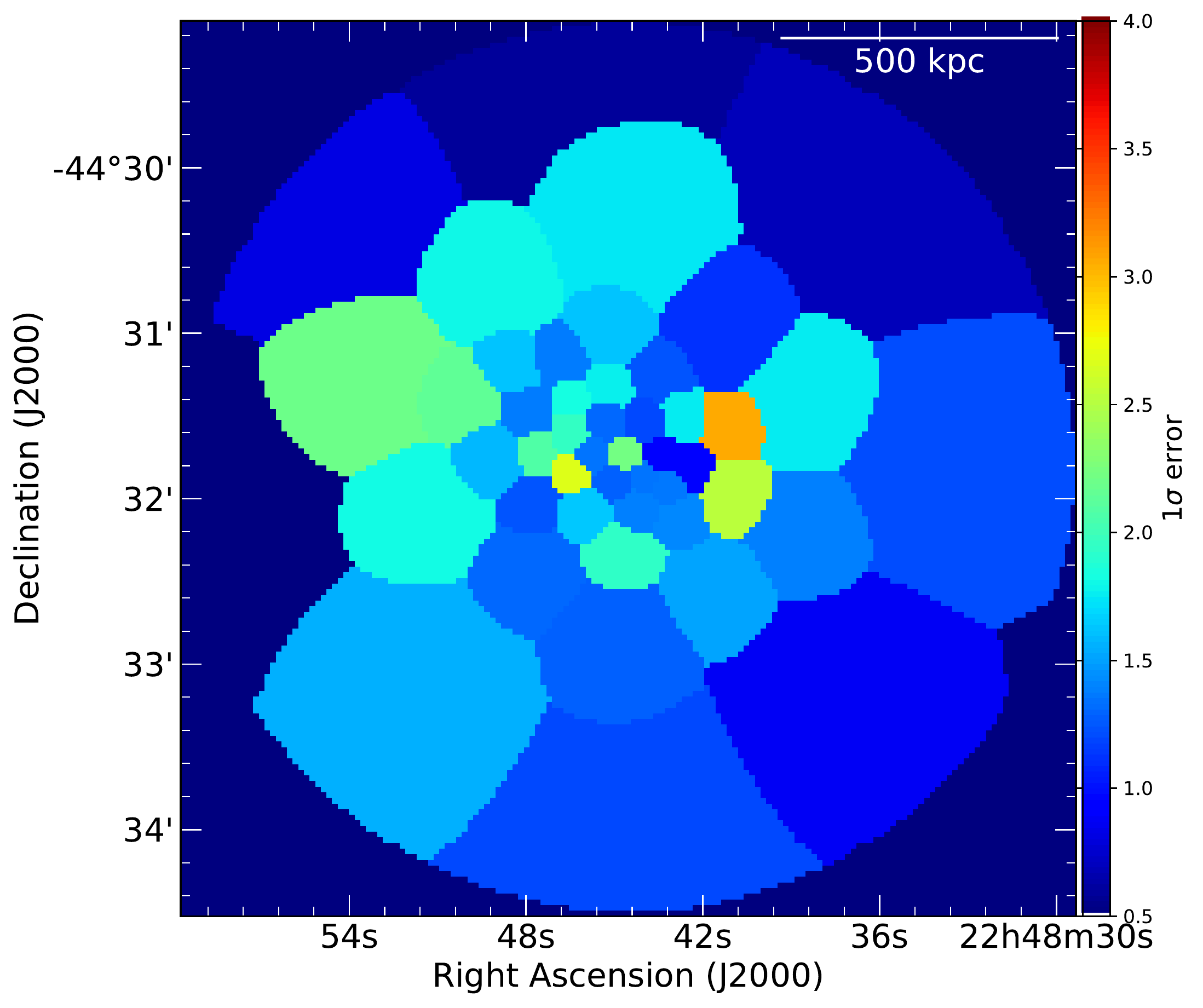} \\
\end{tabular}
\caption{Top: \textit{Chandra} X-ray WVT temperature map of AS1063 in the 0.7-8 keV energy band.
Bottom: 1$\sigma$ error map with respect to the temperature map in the top panel. These images were used to confirm the temperature structure found in the ACB temperature map of the left panel in Figure \ref{fig:ACB_Tmap}.}
\label{fig:WVT}
\end{figure}

In order to confirm the high temperature regions in the ACB map of the left panel in Figure \ref{fig:ACB_Tmap},
we used an alternate temperature map making technique, the Weighted Voronoi Tessellation (WVT) binning algorithm of \citet{Diehl2006}. This WVT-binning algorithm creates non-overlapping regions based on a given SNR. Here, SNR = 50 was used to create the WVT temperature map.
Spectral analysis was performed similarly to that for the ACB method described in Section \ref{sec:TMAP}.
The temperature map obtained using the WVT binning algorithm is shown in the top panel of Figure \ref{fig:WVT}.
The overall temperature structure and the locations of the high temperature regions are consistent with those of the ACB temperature map shown in the left panel of Figure \ref{fig:ACB_Tmap}. This analysis further supports the veracity of the temperature fluctuations found in the ICM of AS1063.

Each of the two temperature map-making techniques has its own advantage and drawback, summarized as follows 
\begin{itemize}
    \item ACB: It allows us to obtain higher resolution temperature structures. However, the temperatures are highly correlated between binned regions.
    
    \item WVT: The temperature information is from spatially independent spectral regions. Hence, temperatures from adjacent regions are not correlated. However, we lose resolution in the resultant temperature map.
\end{itemize}

%%%%%%%%%%%%%%%%%%%%%%%%%%%%%%%%%%%%%%%%% WVT Tmap %%%%%%%%%%%%%%%%%%%%%

%%%% Table 
\begin{table*}
    \begin{center}
    \begin{tabular}{lccccr}
        \hline
        Region & Radius & Abundance & Temperature & Temperature$^a$ & Difference\\
        No. & (arcmin) & (Z$\odot$) & (keV) & (keV) & (In \%) \\
        & & & (using Z from column 3) & (using Z = 0.3Z$\odot$) & \\
        \hline
        1 & 0.00-0.50 & 0.36 & $10.41^{+0.42}_{-0.41}$ & $10.52^{+0.37}_{-0.26}$ & 1.0 \\
        2 & 0.50-1.00 & 0.35 & $12.0^{+0.73}_{-0.74}$ & $12.1^{+0.47}_{-0.39}$ & 0.8 \\
        3 & 1.00-1.50 & 0.28 & $10.64^{+1.15}_{-0.67}$ & $10.62^{+0.80}_{-0.38}$ & 0.2 \\
        4 & 1.50-2.00 & 0.17 & $10.23^{+1.42}_{-1.04}$ & $10.13^{+1.0}_{-0.53}$ & 1.0 \\
        5 & 2.00-3.50 & 0.17 & $9.65^{+1.26}_{-1.18}$ & $9.66^{+1.24}_{-1.17}$ & 0.01 \\
        \hline
    \end{tabular}
    \caption{Abundance table profile}
    \end{center}
    \label{tab:abundance_table}
\end{table*}

%%%%%%%%%%%%%%%%%%%%%%%%%%%% Box region %%%%%%%%%%%%%%%%%%%%%%%%%%%%%%%%%
\begin{figure*}
\centering
\begin{tabular}{c|c}
    \includegraphics[width=3.2in]{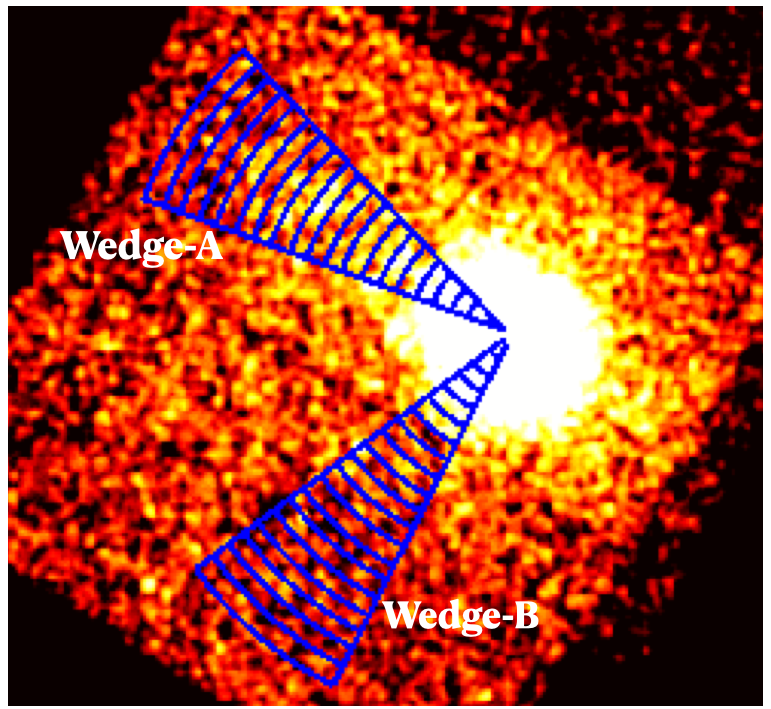} & \includegraphics[width=\columnwidth]{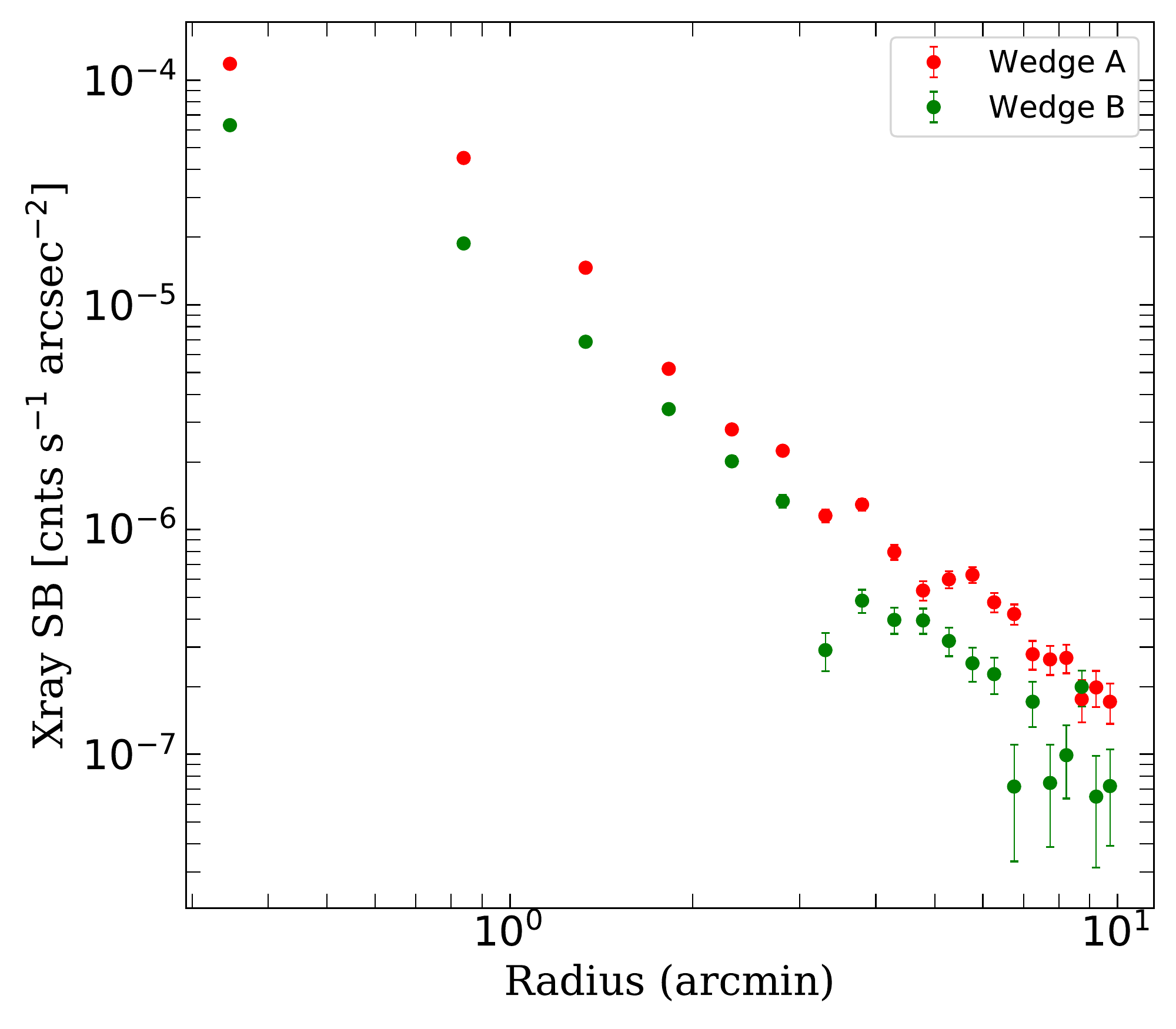} \\
\end{tabular}
\caption{{Left: \textit{Chandra} X-ray surface brightness map of AS1063, overlaid with two wedge shaped regions. Right: \textit{Chandra} X-ray surface brightness profiles over the Wedge-A and Wedge-B regions. }}
\label{fig:Box_a_b}
\end{figure*}
%%%%%%%%%%%%%%%%%%%%%%%%%%%%%%%%%%%%%%%%%%%%%%%%%%
%%%%%%%%%%%%%%%%%%%%%%%%%%%% ACB vs WVT %%%%%%%%%%%%%%%%%%%%%%%%%%%%%%%%%
\begin{figure}
\centering
\includegraphics[width=\columnwidth]{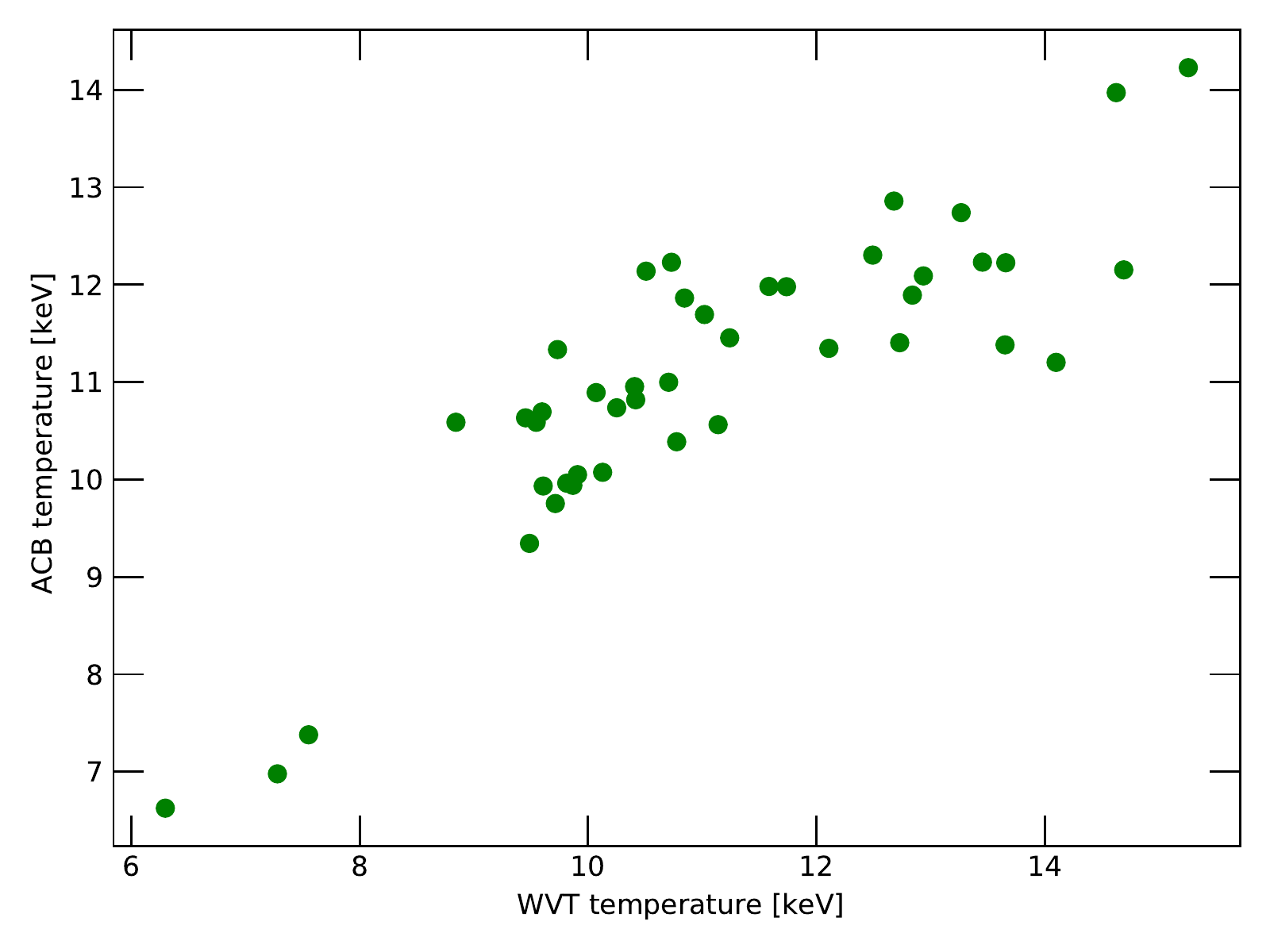}
\caption{Best-fitted temperatures from the ACB and WVT maps. The correlation coefficient between the temperature values of the two maps is found to be 0.85.}
\label{fig:temp_correlation}
\end{figure}

% Don't change these lines
\bsp	% typesetting comment
\label{lastpage}
\end{document}